\documentclass[%
reprint,
twocolumn,
superscriptaddress,
preprintnumbers,
nobibnotes,
nofootinbib,
 amsmath,amssymb, 
 aps, prl,
 longbibliography,
]{revtex4-1}

\usepackage{cancel}
\usepackage{accents}
\usepackage{mciteplus,slashed}
\usepackage{amssymb,cancel,amsmath}
\usepackage{dcolumn}
\usepackage{bm}
\usepackage{soul}
\usepackage[caption=false]{subfig}
\usepackage{appendix}
\usepackage{physics}
\usepackage{booktabs}
\usepackage{comment}
\usepackage[T1]{fontenc}	
\usepackage{csvsimple}
\usepackage[bookmarksnumbered=true]{hyperref} 

\usepackage[dvipsnames]{xcolor}
\definecolor{cernblue}{RGB}{0, 51, 160}
\hypersetup{
    colorlinks = true,
    linkcolor = cernblue,
    anchorcolor = cernblue,
    citecolor = cernblue,
    filecolor = cernblue,
    urlcolor = cernblue
    }
\usepackage[section]{placeins}
\usepackage[capitalise]{cleveref}
\usepackage{booktabs}
\usepackage{graphicx}
\usepackage{mathrsfs}
\graphicspath{{Figures/}}
\AtBeginDocument{
\heavyrulewidth=.08em
\lightrulewidth=.05em
\cmidrulewidth=.03em
\belowrulesep=.65ex
\belowbottomsep=0pt
\aboverulesep=.4ex
\abovetopsep=0pt
\cmidrulesep=\doublerulesep
\cmidrulekern=.5em
\defaultaddspace=.5em
}
\usepackage[dvipsnames]{xcolor}
\usepackage[normalem]{ulem}
\usepackage{fontawesome} 
\usepackage[force]{feynmp-auto}
\usepackage{bbm}
\def\iu{\mathrm{i}}
\def\e{\mathrm{e}}

\usepackage{tikz}
\usetikzlibrary{"arrows", "automata", "backgrounds", "calendar", "chains", "matrix", "mindmap", "patterns", "petri", "shadows", "shapes.geometric", "shapes.misc", "spy", "trees"}
\usetikzlibrary{arrows,shapes}
\usetikzlibrary{decorations.pathmorphing} 
\usetikzlibrary{trees}
\usetikzlibrary{matrix,arrows} 	
\usetikzlibrary{positioning}	
\usetikzlibrary{calc,through}
\usetikzlibrary{decorations.pathreplacing}  
\usepackage{pgffor}	
\usepackage{comment}
\usetikzlibrary{decorations.pathmorphing}	
\usetikzlibrary{decorations.markings}
\tikzset{
    vector/.style={decorate, decoration={snake}, draw},
    vector1/.style={decorate, decoration={snake,amplitude=150pt,segment length=500pt}, draw},
    vector2/.style={decorate, decoration={snake,amplitude=50pt,segment length=250pt}, draw},
	provector/.style={decorate, decoration={snake,amplitude=2.5pt}, draw},
	antivector/.style={decorate, decoration={snake,amplitude=-2.5pt}, draw},
    fermion/.style={draw=black, postaction={decorate},
        decoration={markings,mark=at position .55 with {\arrow[draw=black]{>}}}},
    fermioncyan/.style={draw=black, postaction={decorate},
        decoration={markings,mark=at position .55 with {\arrow[draw=cyan]{<}}}},
    fermiondif/.style={draw=black, postaction={decorate},
        decoration={markings,mark=at position .7 with {\arrow[draw=black]{>}}}},
    fermionend/.style={draw=black, postaction={decorate},
        decoration={markings,mark=at position 1 with {\arrow[draw=black]{>}}}},
    fermionuchannel2/.style={draw=black, postaction={decorate},
        decoration={markings,mark=at position .4 with {\arrow[draw=black]{>}}}},
    scalardif/.style={dashed,draw=black, postaction={decorate},
        decoration={markings,mark=at position .7 with {\arrow[draw=black]{>}}}},
    scalarend/.style={dashed,draw=black, postaction={decorate},
        decoration={markings,mark=at position 1 with {\arrow[draw=black]{>}}}},
    fermionbar/.style={draw=black, postaction={decorate},
        decoration={markings,mark=at position .55 with {\arrow[draw=black]{<}}}},
    fermionnoarrow/.style={draw=black},
    gluon/.style={decorate, draw=black,
        decoration={coil,amplitude=4pt, segment length=5pt}},
    scalar/.style={dashed,draw=black, postaction={decorate},
        decoration={markings,mark=at position .55 with {\arrow[draw=black]{>}}}},
    scalarcyan/.style={dashed,draw=black, postaction={decorate},
        decoration={markings,mark=at position .55 with {\arrow[draw=cyan]{>}}}},
    scalaruchannel1/.style={dashed,draw=black, postaction={decorate},
        decoration={markings,mark=at position .7 with {\arrow[draw=black]{>}}}},
                  scalaruchannel2/.style={dashed,draw=black, postaction={decorate},
        decoration={markings,mark=at position .4 with {\arrow[draw=black]{>}}}},
    scalarbar/.style={dashed,draw=black, postaction={decorate},
        decoration={markings,mark=at position .55 with {\arrow[draw=black]{<}}}},
    scalarnoarrow/.style={dashed,draw=black},
    electron/.style={draw=black, postaction={decorate},
        decoration={markings,mark=at position .55 with {\arrow[draw=black]{>}}}},
	bigvector/.style={decorate, decoration={snake,amplitude=4pt}, draw},
}
\tikzstyle{block} = [draw, rectangle, 
    minimum height=3em, minimum width=6em]
\tikzset{
    cross/.pic = {
    \draw[rotate = 45] (-#1,0) -- (#1,0);
    \draw[rotate = 45] (0,-#1) -- (0, #1);
    }
}
\tikzstyle{block} = [draw, rectangle, 
    minimum height=3em, minimum width=6em]

\definecolor{amber}{rgb}{1.0, 0.75, 0.0}
\definecolor{darkturquoise}{rgb}{0.0, 0.81, 0.82}
\definecolor{mediumaquamarine}{rgb}{0.4, 0.8, 0.67}
\definecolor{coralred}{rgb}{1.0, 0.25, 0.25}

\newcommand{\bsq}{\,\rule[0pt]{3.5pt}{3.5pt}\,}
\newcommand\cone{\ensuremath{{\color{amber}\bullet}}}
\newcommand\ctwo{\ensuremath{{\color{mediumaquamarine}\bsq}}}
\newcommand\cthree{\ensuremath{{\color{violet}\blacktriangle}}}

\newcommand{\FF}{G(\vb{q})}
\newcommand{\cosqx}{\cos(\vb{q} \cdot \vb{\Delta x})}
\newcommand{\sinqx}{\sin(\vb{q} \cdot \vb{\Delta x})}
\newcommand{\qu}{L}
\newcommand{\TT}{\mathcal{T}}
\newcommand{\Y}{\mathcal{Y}}

\definecolor{interorange}{RGB}{1.0,0.3098,0}
\definecolor{forestgreen}{HTML}{228B22}

\begin{document}

\title{Coherently enhanced decoherence and cloud substructure of atom interferometers}

\author{Clara Murgui}
\email{clara.murgui@cern.ch}
\affiliation{Theoretical Physics Department, CERN, 1 Esplanade des Particules, CH-1211 Geneva 23, Switzerland}
\author{Ryan Plestid}
\email{ryan.plestid@cern.ch}
\affiliation{Walter Burke Institute for Theoretical Physics, California Institute of Technology, Pasadena, CA 91125, USA}
\affiliation{Theoretical Physics Department, CERN, 1 Esplanade des Particules, CH-1211 Geneva 23, Switzerland}

\date{\today}

\preprint{CALT-TH/2025-028~~~CERN-TH-2025-174}

\begin{abstract}
  We study how coherent scattering of a background gas off an atom (or other matter) interferometer can lead to enhanced signals from phase shifts and contrast loss. We focus on the inclusion of realistic features of atom interferometers such as finite temperature, cloud substructure, and time-dependent cloud radii. The inclusion of these effects, extending beyond the previously considered point-like cloud approximation,  naturally allow us to study the smooth transition between the coherent and incoherent scattering regimes. We discuss how the formalism presented herein can be tested in the laboratory (with near-infrared photons or an eV-scale electron gun), and discuss an application for the detection of dark matter interacting via long-range forces. 
  
\end{abstract}

 \maketitle 

~ 

\vspace{-16pt}

\section{Introduction}
The intricate control of quantum systems has emerged as a key tool for investigating fundamental quantum physics. Beyond issues in quantum foundations, there has recently been a groundswell of interest in quantum sensing for problems related to issues in gravitational and particle physics. 

Atom interferometers were conceived as accelerometers and sensors for measuring some of the fundamental constants of nature~\cite{Kasevich:1992yii,Zhou:2015pna,Rosi:2017ieh,Yu:2019gdh,Morel:2020dww}. They have emerged as a promising platform to search for ultra-light dark matter \cite{Arvanitaki:2016fyj,Geraci:2016fva,Blas:2024kps,Badurina:2025xwl}, gravitational waves \cite{Dimopoulos:2007cj,Graham:2012sy,Graham:2016plp}, fifth forces~\cite{Wacker:2009ag,Graham:2015ifn,Abe:2024idx}, among other phenomena beyond the Standard Model. Of particular note is the proposal to search for phase shifts and/or a loss of contrast induced by a background gas of virialized dark matter particles~\cite{Riedel:2012ur,Riedel:2016acj,Du:2022ceh}. This phenomenon lies beyond the often-considered classical-field limit for ultralight dark matter, and is known as collisional decoherence~\cite{Joos:1984uk,PhysRevA.42.38,Hornberger_2003,Tegmark:1993yn,Giulini:1996nw}, and offers a threshold-free detection mechanism, which lies in stark contrast to conventional particle physics detectors which are blind to depositions below their energy thresholds (see, for example, \cite{Zurek:2024qfm}).

In recent work~\cite{Badurina:2024nge}, we have developed an extension of the single-atom framework for collisional decoherence~\cite{Hornberger_2003} that accounts for macroscopic coherent enhancements from an ensemble of atoms. Coherent effects can increase certain observables by orders of magnitude, concretely by the square of the number of atoms that interact coherently with the background gas. Since atom interferometers involve clouds of millions of atoms, it is crucial to have reliable theory of these coherent effects in order to accurately interpret interferometer data, particularly when searching for low momentum transfer scattering of elusive particles (e.g., dark matter) with atoms.

While Ref.~\cite{Badurina:2024nge} put forth a formalism for calculating decoherence rates, phenomenological applications were restricted to the point-like limit (where the cloud radius tends to zero). This approach ignored experimentally realistic issues such as the finite temperature of the cloud, its density profile, and by extension its time-dependent radius. The cloud's density  profile is a key input in the theory of coherent scattering since its Fourier transform yields the elastic form factor \cite{Riedel:2016acj,Bednyakov:2018mjd}.

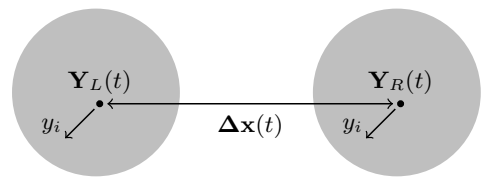
\begin{figure}[t]
\begin{equation*}
\begin{gathered}
\begin{tikzpicture}[line width=1.5 pt,node distance=1 cm and 1 cm]
\filldraw[color=lightgray] (-0.05,0.15) circle (31pt);
\filldraw[color=lightgray] (3.95,0.15) circle (31pt);
\coordinate[label=$\vb{Y}_L(t)$](xg);
\coordinate[below left = 0.75 cm of xg,label=$y_i \ \ $](ygi);
\coordinate[right = 4 cm of xg,label=$\vb{Y}_R(t)$](xe);
\coordinate[left = 0.1 cm of xe](xeaux);
\coordinate[below left = 0.1 cm of xg](xgaux2);
\coordinate[below left = 0.1 cm of xe](xeaux2);
\coordinate[above right = 0.1 cm of ygi](ygiaux);
\coordinate[right = 0.1 cm of xg](xgaux);
\coordinate[right = 2 cm of xg,label=below:$\vb{\Delta x}(t)$](Deltax);
\coordinate[right = 4 cm of ygi,label=:$y_i \ \ $](yei);
\coordinate[above right = 0.1 cm of yei](yeiaux);
\draw[fill=black] (xg) circle (.02cm);
\draw[fill=black] (xe) circle (.02cm);
\draw [line width=0.2mm,<->] (xgaux) -- (xeaux);
\draw [line width=0.2mm,->] (xgaux2) -- (ygiaux);
\draw [line width=0.2mm,->] (xeaux2) -- (yeiaux);
\end{tikzpicture}
\end{gathered}
\end{equation*}
\caption{Schematic of an atom interferometer. Two spatially extended clouds of atoms are placed into a quantum superposition of excited and ground states. The momentum imparted by the pulse sequence causes atoms in the excited state to become spatially separated from atoms in the ground state by a distance $\vb{\Delta x}(t)$. We define the atoms' coordinates $\vb{x}_i(t=0) = \vb{y}_i$ relative to the center of mass of the cloud. The classical trajectories,  $\vb{Y}_{L,R}(t)$, are different for ground and excited states and lead to macroscopic separations $\vb{\Delta x}(t)$ at intermediate times. \label{fig:clouds} }
\end{figure}

In this paper we extend the results of Ref.~\cite{Badurina:2024nge} to account for the above-mentioned physical effects. A schematic of the setup we consider is shown in \cref{fig:clouds}.  We organize our discussion around two distinct approximation schemes: one which neglects cloud-spreading but treats decoherence exactly, and one which includes cloud-spreading but treats decoherence perturbatively. We apply our results to a thermal atom cloud prepared in an harmonic trap, provide explicit computations of the induced phase shift for an infrared photon lamp and $1~{\rm eV}$ electron gun, and comment on the application of our formalism to searches for dark matter.

\vfill 
\pagebreak

\section{Collisional decoherence
and atom interferometers \label{sec:background} }
We are interested in atom interferometers, and specialize to a two-arm setup for concreteness. We focus only on ``center of mass'' interferometry in which atoms inside a cloud are placed into a coherent superposition of states that trace macroscopically separated paths.

Atom interferometers can be influenced by scattering with a rarefied background gas. 
This phenomenon is known as ``collisional decoherence'' \cite{Gallis:1990,Hornberger:2003}, and can be dramatically enhanced when the background gas scatters coherently off the entirety of the atom cloud (containing $N\gg 1$ atoms).

Tracing over the density matrix of the background gas and neglecting all atomic recoil in the collisions, the rate of change due to collisional decoherence is given by \cite{Gallis:1990,Hornberger_2003,Badurina:2024nge}
\begin{equation}
    \label{EOM-coll}
  \qty(\pdv{\rho}{t})_{\rm coll.} =  - \qty(\lambda_U + \lambda_D)\rho \, ,
\end{equation} 
where we have written our result in the position representation, and in the basis of the internal states of the atoms, $\{s\}$, i.e.
\begin{equation}
    \rho= \rho(\{\vb{x}\},\{\vb{x}'\}; \{s\},\{s'\}; t)    \, .
\end{equation}
Without loss of generality, we take these to be the ground and an excited state of the atom.\!\footnote{The excited state has to be sufficiently long-lived to last over the experiment. Typically, it corresponds to some hyperfine splitting of the atomic ground state \cite{Cronin:2009zz}.} These internal states are used to generate interferometer paths; however, during the interferometer's pulse sequence, the mapping between the internal states $\{e,g\}$ and the two paths $\{L,R\}$ changes. Nevertheless, the mapping remains bijective at the read-out time $\mathcal{T}$.

Including the rate of change due to the interferometer's own isolated dynamics, assumed to be generated by a Hamiltonian $H_{\rm sys.}$, the time evolution of the density matrix is given by
\begin{equation}
  \label{full-EOM}
  \pdv{\rho}{t} = -\iu \qty[ H_{\rm sys.}, \rho] - \qty(\lambda_U + \lambda_D)\rho~, 
\end{equation} 
where we work in the position representation. In what follows, we take $H_{\rm sys.}$ to the be the free Hamiltonian 
\begin{equation}
    \label{Hsys}
    H_{\rm sys.} = \sum_{i=1}^N -\frac{\vb*{\nabla}_i^2}{2M}~, 
\end{equation}
where $M$ is the mass of the atom. \Cref{full-EOM} can be written in Lindblad form where $\lambda_D$ is generated by jump-operators~\cite{Badurina:2024nge}. 

The right-hand side of \cref{EOM-coll} contains a ``unitary'' ($\lambda_U$) and ``decohering'' ($\lambda_D$) piece, induced by interactions with probe particles from a dilute, background gas.\footnote{We assume the background to be homogeneous and Markovian.}
These are as integrals over kernels~\cite{Badurina:2024nge}
\begin{align}
    \lambda_U &= \int_{\qu}  \omega_U(\vb{\qu}) \,  K_U(\vb{L},\{\vb{x}\},\{\vb{x}'\})~,\label{eq:lambdaU}\\
    \lambda_D &= \int_{q}  \omega_D(\vb{q}) \, K_D(\vb{q},\{\vb{x}\},\{\vb{x}'\})~,\label{eq:lambdaD}
\end{align}
where $\int_{k} \equiv \int \dd^3 k /(2\pi)^3$, while the leading Born-series (perturbative) expressions for $\omega_{U,D}$ and $K_{U,D}$ are given explicitly below. 

Consider an interaction potential\footnote{In relativistic applications $\tilde{V}(\vb{q})$ is replaced by a relativistic matrix element \cite{Badurina:2024nge}.} in momentum space, $\hat{V}(\vb{q})\sim O(g)$ with $g\ll 1$ a perturbative coupling quantifying the interaction between the probe and the atom. Then, to second order in the Born series, the real-valued functions $\omega_{U,D}$ for non-relativistic scattering are given by 
\begin{align}
\omega_U(\vb{\qu}) &= \int_p \!\! \rho_\pi (\vb{p})  |\tilde V(\vb{\qu})|^2 2 \text{Re}[G_\pi(\vb{p}+\vb{\qu})]~, \\
\omega_D(\vb{q}) &= \int_p \rho_\pi(\vb{p})  (2\pi) \delta ( \Sigma E)|\tilde V(\vb{q})|^2~,
\end{align}
where $G_\pi$ and $\rho_\pi$ are the probe propagator and  density matrix, respectively,  and  $\delta (\Sigma E)$ enforces energy conservation. The kernels $K_{U,D}$ are given by
\begin{align}
    \!\!K_U & =  \frac{\iu}{2} \sum_{ij}^N\e^{\iu \vb{L}\cdot (\vb{x}_i' - \vb{x}_j')}-e^{\iu \vb{L}\cdot (\vb{x}_i-\vb{x}_j)}~, \\
    \!\!K_D& = \frac{1}{2}\sum_{ij}^N \e^{\iu \vb{q}\cdot (\vb{x}_i - \vb{x}_j)} +\e^{\iu \vb{q}\cdot (\vb{x}_i'-\vb{x}_j')} -2\e^{\iu \vb{q}\cdot (\vb{x}_i-\vb{x}_j')}~.
\end{align}
The above expressions keep only the leading contributions to decoherence, proportional to $g^2$.

Since we specialize to two-arm interferometers, it will often be convenient to define a set of ``comoving'' coordinates $\vb{y}_i$ which account for the displacements of the atoms with respect to the cloud's center of mass, $\vb{Y}_{\mathcal{P}_i}(t)$,
\begin{equation}
    \label{comoving-coords}
    \vb{x}_i = \vb{Y}_{ \mathcal{P}_i}(t) + \vb{y}_i \, , 
\end{equation}
where $\mathcal{P}_i$ is the path taken by the $i^{\rm th}$ atom, as dictated by its internal label $s_i$. Note that, for a superposition, this means that the two branches of the atom's wavefunction are described by two different sets of comoving coordinates.  See \cref{fig:clouds} for a schematic illustration of their meaning.

The trajectory $\vb{Y}_{ \mathcal{P}_i}(t)$ solves the classical equations of motion subject to initial conditions, with initial momentum $\vb{p}_{0} = \vb{0}$ for the ground state, and $\vb{p}_0 = \vb{k}_\gamma$ for the excited state, where $\vb{k}_\gamma$ is the momentum imparted by the pulse sequence (which can be much larger than the level splitting \cite{Rudolph:2019vcv,AION:2025igp}). At a later time, a $\pi$-pulse is applied and the internal states are exchanged. The trajectories then are ``kicked'', leading to convergence of the clouds at the time of measurement, $t=\mathcal{T}$, such that the trajectories  coincide, $Y_{L}(\mathcal{T})=Y_R(\mathcal{T})$. Finally, a $\pi/2$-pulse is applied to decouple the trajectories from the internal degrees of freedom, which are then used for readout.\footnote{We have in mind a Mach-Zender $\pi/2$-$\pi$-$\pi/2$ pulse sequence. While we are aware that atom interferometer architectures differ in their technical implementation, our discussion applies equally well to any system with entanglement between the center of mass and internal degrees of freedom.} 

When defining our initial conditions, we consider a (pure or mixed) quantum state, $\varrho_0$. We define $\varrho_0$ such that its center of mass does not evolve in time, and its dynamics lead to trivial interferometry (i.e., it just ``sits there''). Next, some state preparation is performed, 
\begin{equation}\label{state-prep}
  \varrho_0 \rightarrow \begin{gathered} \boxed{\underset{\text{e.g. }\pi/2 \text{ pulse}}{\text{\scriptsize state  prep.}}} \end{gathered} \rightarrow \rho_0,
\end{equation}
such that $\rho_0$ now has a {\it non-trivial} evolution in the interferometer.

As a concrete example, let us consider a cloud of atoms in a trap. The atoms will, in general, have some distribution $\varrho_0$ which is localized in some region of space, for example a thermal state in a harmonic trap. However, we will assume that all atoms are initially in their own internal ground states, $\ket{g}$, 
\begin{equation}
    \varrho_0 = \varrho_y(\{\vb{y}\},\{\vb{y}'\}) \otimes \varrho_s(\{s\},\{s'\})~,
\end{equation}
where $\varrho_s(\{s\},\{s'\}) = \bigotimes_i \ket{g_i}\bra{g_i}$. At $t=0$, coordinates $\vb{y}$ and $\vb{x}$ coincide and we have $\vb{y}_i=\vb{x}_i$. Next, a $\pi/2$-pulse is applied to all of the atoms, placing them in a superposition of ground and excited states. %
For an atom-cloud we then have, at $t=0$, 
\begin{equation}\label{rho0}
  \rho_{0} = \varrho_y(\{\vb{y}\},\{\vb{y}'\}){ \, \otimes \, \textstyle  \bigotimes_i }\big(\ket{\psi_i} \bra{\psi_i}\big)~,
\end{equation}
where $ \ket{\psi_i}=  \tfrac1{\sqrt{2}}(\ket{g_i} + \e^{\iu \vb{k}_\gamma\cdot \vb{y}_i}\ket{e_i})$.  Notice that the state $\ket{\psi_i}$ has a non-trivial dependence on $\vb{y}_i$, and the coordinate-space and internal degrees of freedom have become entangled.

The clock starts. The imparted momentum $\vb{k}_\gamma$ now generates a separate trajectory for the $\ket{e}$ state vs. the $\ket{g}$ state. The state of the system then bifurcates into many branches upon evolution under the unitary dynamics, induced by $H_{\rm sys.}$. The state $\rho_0$ in \cref{rho0} then serves as an initial condition for \cref{full-EOM}, and leads to a time-dependent density matrix $\rho(t)$. During time evolution, the environment induces non-unitary decoherence. This phenomenon is described in detail for single atoms in Refs.~\cite{Gallis:1990,Hornberger:2003}, and for point-like clouds with $N\geq 2$ atoms in Ref.~\cite{Badurina:2024nge}. 

As discussed above, measurements are performed at some time $t=\mathcal{T}$, defined such that, after a certain combination of laser pulses, (a) the two trajectories coincide,  i.e. $\vb{Y}_L(\mathcal{T}) = \vb{Y}_R(\mathcal{T})$, and (b) the coordinate-space and internal degrees of freedom disentangle. These guarantee that the measurement will not projectively decohere the interferometer, as it would if performed at intermediate times when internal and center of mass degrees of freedom are still entangled and the two interferometer paths are spatially separated. 

Next, one must define the observable of interest. Often it only requires some partial information about the matter-wave's density matrix, and one can trace over all other degrees of freedom such as the positions of the atoms. Typically, to detect the atom number of the interferometer outputs, only the internal state of the atom is read-out and the position degrees of freedom can be traced over. This read-out is done after a pulse sequence and physically the $\ket{g}$ and $\ket{e}$ states of the atom are directly interrogated (e.g. via fluorescence~\cite{RevModPhys.81.1051,Rocco_2014}). However, from a theoretical perspective, it is convenient to work in the basis of states prior to the final pulse applied, such that the experimentalist effectively interrogates two states which we will take to be $\ket{\pm} = \frac{1}{\sqrt{2}} \qty(\ket{g} \pm\ket{e})$.

If one counts the number of atoms in the ``$+$'' port, then the relevant operator may be written for the $i^{\rm th}$ atom as
\begin{equation}
    \label{O-plus}
    \begin{split}
    \frac{1}{\mathcal{V} }\int \dd^3x_i \big(\ket{\vb{x}_i}\bra{\vb{x}_i} \big) \otimes \mathcal{O}_i~,
    \end{split}
\end{equation}
with $\mathcal{O}_i=\big(\ket{+}\bra{+}\big)_i$, where we have normalized by the volume of space, $\mathcal{V}$, over which we integrate. A counting experiment corresponds to measuring
\begin{equation}
    \mathcal{O}_+ = \sum_{i=1}^N \mathcal{O}_i~. 
\end{equation}
Other observables can be constructed by raising $\mathcal{O}_+$ to higher powers. These are related to higher statistical moments; e.g., $\langle \mathcal{O}_+^2\rangle$ is related to the variance.

The form of \cref{O-plus} makes it clear that one can trace over all of the atomic coordinates to compute the expectation value of the observable for a system of $N$ atoms,
\begin{equation}
    \begin{split}
    \langle \mathcal{O}_+ \rangle_{t=\mathcal{T}} &= {\rm Tr}\qty[\mathcal{O}_+ \rho(\mathcal{T})  ] ~\\
    &={\rm Tr}_{\{s\}}\qty[\mathcal{O}_+ {\rm Tr}_{\{\vb{x}\}}\qty[\rho(\mathcal{T})]  ]~. 
    \end{split}
\end{equation}
It is therefore convenient to introduce the $2^N\times 2^N$ reduced density matrix, $\rho_r(\{s\},\{s'\}) = {\rm Tr}_{\{\vb{x}\}}\qty[ \rho(\mathcal{T})]$, written only in the basis of internal degrees of freedom $\{s\}$ and $\{s'\}$.

Conventional atom interferometers measure a fringe built from the number of counts, $N_\pm$,  in the $(\pm)$-ports defined above. The relative population $N_+/(N_++N_-)$ is plotted against different choices of a tuneable phase-shift, $\Phi$, between the two paths of the interferometer. The tuneable phase-shift $\Phi$ depends on experimental parameters and the energy splitting between $\ket{g}$ and $\ket{e}$.

Atom interferometer data is composed of a series of data points, $\{ n_+^{(1)}, n_+^{(2)}, \cdots, n_+^{(\aleph)}\}$, obtained after projecting the internal state of the atoms. The counting index $\aleph$ corresponds to the number of times the experiment is repeated for a given $\Phi$. The number $n_\pm^{(i)}$ is how many atoms were counted to be in the $\ket{\pm}$ state in the $i^{\rm th}$ iteration of the experiment.

Denoting experimental quantities with a subscript ``exp'', we define
\begin{equation}
  \left\langle \frac{N_+}{N_++N_-} \right\rangle_{\rm exp}   \equiv \frac{1}{\aleph} \sum_{i=1}^\aleph \frac{n_+^{(i)}}{n_+^{(i)}+n_-^{(i)}}~.
\end{equation}
The measured visibility or contrast, $V$, and phase-shift, $\varphi$, are given by fitting a sinusoidal curve to the observable, 
\begin{equation}
    \label{best-fit-V-phi}
       \left\langle \frac{N_+}{N_++N_-} \right\rangle_{\rm exp}~~  \underset{\text{\tiny best fit}}{\longrightarrow} ~~\frac{1}{2} \bigg(1+V\cos(\Phi+\varphi)\bigg)~,
\end{equation}
for different values of $\Phi$ (which we assume to be known). We note that both the left- and right-hand side depend on the tuneable phase $\Phi$.

The experimentally determined quantities will converge (in a frequentist sense) to the theory prediction
\begin{equation}\label{eq:expth}
\left\langle \frac{N_+}{N_++N_-} \right\rangle_{\rm th} \equiv \frac{1}{N} \langle \mathcal{O}_+ \rangle = \text{Tr} \big[~\rho_1 \ket{+}\bra{+}~\big]~,
\end{equation} 
where $\rho_1$ is the one-body density matrix, $\rho_1 \equiv \text{Tr}_{N-1}\qty[\rho_r]$.

In the absence of interactions of the environment, \cref{eq:expth} gives $V=1$ and $\varphi=0$. Deviations from these expectations can arise from unitary effects, which induce a phase shift, $\varphi \neq 0$. A loss of contrast can arise from multiple sources: (1) genuine decoherence in which off-diagonal entries of $\rho_r$ decrease in magnitude, (2) dephasing in which an average over unitary phases results in destructive interference, or (3) classical stochastic processes such as a time-varying phase shift which is ``averaged out'' in the data. In what follows, we will only discuss scenarios (1) and (2).

Having defined the state preparation protocol, the equations of motion which govern time evolution, and the observable(s) of interest, we now turn to explicit solutions for $\rho(t)$. 

Most atom interferometers satisfy the hierarchy 
\begin{equation}
    \label{real-hierarchy}
    t_{\rm kick} \ll t_{\rm spread} \ll t_{\rm decoh.}~.
\end{equation}
Here,  $t_{\rm kick}$ is defined as fothe time taken from the clouds to be separated by more than one cloud radius, $t_{\rm spread}$ is the time scale over which the cloud's radius changes by an $O(1)$ amount, and $t_{\rm decoh.}$ is the time scale over which the interferometer decoheres by one $e$-fold.

In what follows, we will first examine a hierarchy different than \cref{real-hierarchy}, in which the cloud's structure changes very slowly in time. In this limit, one can employ a ``static approximation'', in which the spreading of the cloud is neglected, while collisional decoherence persists. This is a convenient analytical limit that provides clear insight into the dynamics of an atom interferometer. After completing this analysis, we return to the hierarchy in \cref{real-hierarchy} and discuss the relevant dynamics in \cref{sec:naive-TDPT}. 

\vfill

\section{Static approximation \label{sec:static} }
%
It is convenient to consider the limit in which cloud spreading is negligible, but decoherence is not. In this limit, the cloud is ``static'' in the sense that it does not spread, but its center of mass moves along the classical trajectories $\vb{Y}_{L,R}(t)$. This limit is tractable, and physically realized whenever
\begin{equation}
    t_{\rm kick} \sim t_{\rm decoh.} \ll t_{\rm spread}~.
\end{equation}
It is therefore well suited to the limit of slowly spreading, or rapidly decohering atom interferometer. This maps cleanly to the point-like limit considered in Ref.~\cite{Badurina:2024nge} (where cloud structure was neglected entirely), and can be applied when decoherence rates are large. Furthermore, the static limit is applicable to matter interferometry (where inter-atomic positions are fixed), which we briefly discuss in \cref{sec:applications}. 

The static limit described above corresponds to a semi-classical approximation where gradients of the density matrix are neglected, but one retains $\vb{k}_\gamma$ from the laser-kick. Mathematically, this corresponds to 
\begin{equation}
  \begin{split}
    -\iu \frac{1}{2M} {\vb*{\nabla}}^2 \e^{\iu \vb{k}_\gamma \cdot \vb{x}} &=
    \e^{\iu \vb{k}_\gamma \cdot \vb{x}}
    \qty(\frac{1}{M}\vb{k}_\gamma\cdot \vb*{\nabla} +  \iu \frac{1}{2M} {\vb*{\nabla}}^2) \\
    & \simeq \e^{\iu \vb{k}_\gamma \cdot \vb{x}} \vb{v}\cdot \nabla~,
    \end{split}
\end{equation}
where $\vb{v}=\vb{k}_\gamma/M$. 

The solution of the equations of motion in \cref{full-EOM} can then be obtained using the method of characteristics. It is worth noting that a background for the atoms, such as the Earth's gravitational field, can be straightforwardly incorporated within the semi-classical approximation by altering the classical trajectories to solve the equations of motion in the presence of a potential, but for simplicity we do not consider this here.

The result is that the solution for an atom interferometer, evolving under the free Hamiltonian in \cref{Hsys}, can be written as 
\begin{widetext}
  \begin{equation}
    \label{eq:expl-soln-semiclassical}
  \rho(\{\vb{x}\},\{\vb{x}'\};\{s\},\{s'\};\mathcal{T}) =  \varrho_y(\{\vb{y}\},\{\vb{y}'\})\varrho_s(\{s\},\{s'\})
  \e^{\iu \tilde{\Phi}(\{s\},\{s'\})}
  \exp\qty[ - \int_0^\mathcal{T} \dd t \,   \lambda(\{\vb{x}_s(t)\},\{\vb{x}'_{s'}(t)\}) ]~,
\end{equation}
\end{widetext}
where $\lambda =\lambda_D+\lambda_U$, and $\tilde{\Phi}(\{s\},\{s'\})$ depends on the internal states of the matrix element. \Cref{eq:expl-soln-semiclassical} resums effects of order $g^2\mathcal{T}$ in the limit of negligible cloud-spreading. 

The coordinates in the exponential are defined in \cref{comoving-coords}, and encode the ``which path'' information via the labels $s$ and $s'$. Notice that the dependence on the initial coordinate-space density matrix, $\varrho_y(\{\vb{y}\},\{\vb{y}'\})$,  {\it factorizes} from the rest of the expression. This non-trivial property follows from the fact that collisional decoherence is diagonal in the coordinate representation, {\it and} from the pulse sequence of lasers which leads to all branches of the wavefunction having zero center-of-mass momentum at the time of measurement; this final property is realized by definition at $t=\mathcal{T}$.

The phase $\tilde{\Phi}(\{s\},\{s'\})$ can be expressed in terms of the single-atom phase $\Phi$, cf. \cref{best-fit-V-phi},
\begin{equation}
    \tilde{\Phi}(\{s\},\{s'\}) = n ~\Phi~,
\end{equation}
with $\Phi$ being the tuneable relative single atom phase-shift between the two atom interferometer arms. The variable $n$ is defined via 
\begin{equation}
    \label{little-n-def}
    n \equiv N_L-N_L'~,
\end{equation}
where $N_L$ is the number of atoms in the left arm of the interferometer in $\ket{\{\vb{x}\}}$, and $N_L'$ is the number of atoms in the left-arm of the interferometer in $\bra{\{\vb{x}'\}}$.

Having constructed an explicit solution for $\rho(\mathcal{T})$, we can now perform the trace over position to obtain 
\begin{equation}
    \rho_r(\{s\},\{s'\}) \equiv {\rm Tr}_{\{\vb{x}\}}\qty[~ \rho(\{\vb{x}\},\{\vb{x}'\};\{s\},\{s'\};\mathcal{T})~]~.
\end{equation}
Defining $\int_y\equiv \frac{1}{\mathcal{V}} \int \dd^3y $, with $\mathcal{V}$ the volume of integration, we have 
\begin{equation}
  \label{decoherence-factor}
 \rho_r \propto \int_y~\varrho_y(\{\vb{y}\},\{\vb{y}\})
  \e^{- \int_0^\mathcal{T} \dd t  \lambda(\{\vb{x}_s(t)\},\{\vb{x}'_{s'}(t)\}) }~.
\end{equation}
We see that $\rho_r$ is indexed by the labels $\{s\}$ and $\{s'\}$ and has a dimensionality of $2^N\times 2^N$.

Within $\rho_r$, the decoherence terms $\lambda_U$ and $\lambda_D$ simplify to
\begin{align}
    \lambda_U &= \iu \int_{\qu}  \omega_U(\vb{\qu})  \sum_{ij}^N \Y_U^{ij}(\vb{\qu},t) \,\e^{i \vb{\qu}\cdot (\vb{y}_i -\vb{y}_j)}~,\label{eq:lambdaU}\\
    \lambda_D &= \int_{q}  \omega_D(\vb{q}) \sum_{ij}^N  \Y_{D}^{ij}(\vb{q},t) \, \e^{\iu\vb{q}\cdot (\vb{y}_i-\vb{y}_j)}~~~.\label{eq:lambdaD}
\end{align}
All time dependence is contained in
\begin{align}
    \Y_{U}^{ij}(\vb{L},t) &= \frac{1}{2}\left[\e^ {\iu \vb{\qu}\cdot(\vb{Y}_{\mathcal{P}_i'}  -  \vb{Y}_{\mathcal{P}_j'})}-\e^{\iu\vb{\qu}\cdot(\vb{Y}_{\mathcal{P}_i}-\vb{Y}_{\mathcal{P}_j})}\right],\\
    \begin{split}
    \Y_{D}^{ij} (\vb{q},t) &= \frac{1}{2}\left[ \e^{\iu \vb{q}\cdot(\vb{Y}_{\mathcal{P}_i}-\vb{Y}_{\mathcal{P}_j})} + \e^ {\iu\vb{q} \cdot (\vb{Y}_{\mathcal{P}_i'} - \vb{Y}_{\mathcal{P}_j'})} \right. \\
    &\left. \hspace{0.34\linewidth}-2 \e^{\iu\vb{q} \cdot(\vb{Y}_{\mathcal{P}_i} - \vb{Y}_{\mathcal{P}_j'}) } \right]~,
    \end{split}
\end{align}
where the classical trajectory variable can only realize two possible outcomes at a given time, either $\vb{Y}_L(t)$ or $\vb{Y}_R(t)$. Using \cref{eq:lambdaD,eq:lambdaU}, we can factor out $\e^{\iu \vb{q} \cdot (\vb{y}_i -\vb{y}_j)}$ (or $\e^{\iu \vb{\qu} \cdot (\vb{y}_i -\vb{y}_j)}$) from the integral over $\dd t$ in the exponential of \cref{decoherence-factor}. The effects of $\e^{\iu \vb{q} \cdot (\vb{y}_i -\vb{y}_j)}$ (or $\e^{\iu \vb{\qu} \cdot (\vb{y}_i -\vb{y}_j)}$) are then realized after performing the partial trace over coordinates $\int_y$. 

It is informative to expand the exponential and consider each term in the Taylor series, 
\begin{equation}
    \e^{-\int_t \lambda(t)} = 1 - \int_t \lambda(t) + \frac12 \int_t \int_{ t' }\lambda(t)\lambda(t') + \ldots ~,
\end{equation}
with $\int_t \equiv \int_0^\mathcal{T} \dd t$, which defines a related expansion, 
\begin{equation}
    \rho_r =  \rho_r^{(0)} +\rho_r^{(1)} + \rho_r^{(2)} + \ldots ~.
\end{equation}
At first non-trivial order (i.e., for $\rho_r^{(1)}$) we encounter the correlation function \cite{Bednyakov:2018mjd} (or the Debye-Waller factor \cite{lipkin2004physicsdebyewallerfactors})
\begin{equation}\label{eq:G}
  G_{ij}(\vb{q}) \equiv  \int_y \varrho_y(\{\vb{y}\},\{\vb{y}\}) \e^{\iu \vb{q} \cdot (\vb{y}_i -\vb{y}_j)} ~,
\end{equation}
which is given by
\begin{equation}
  G_{ij}(\vb{q}) = \begin{cases}
    1         & i=j~,   \\
    G(\vb{q}) & i \neq j~.
    \end{cases} 
\end{equation}
We then obtain
\begin{widetext}
    \begin{equation}
        \begin{split}
        \!\!\!\!\rho_r^{(1)}(\mathcal{T}) = &  \ \e^{\iu n \Phi}\varrho_s(\{s\},\{s'\})\qty[\int_{q,t} \!\!\! \omega_D \bigg [ \big(1-G(\vb{q}) \big) \sum_i \Y_D^{ii}(t,\vb{q}) 
         \!+\! G(\vb{q}) \sum_{ij} \Y_D^{ij}(t,\vb{q})\bigg ]
         \!\!+ \iu\! \int_{\qu,t}\!\!\! \omega_U \ G(\vb{\qu}) \sum_{ij}\Y_U^{ij}(t,\vb{\qu}) ]~.\! \label{eq:lambdaG}
         \end{split}
    \end{equation}
\end{widetext}
This expression captures the incoherent (scales like $\sim N$)  and coherent (scales like $\sim N^2$) contributions to $\lambda_{U,D}$. Using \cref{little-n-def}, the sums (running from $i,j=1$ to $N$) over $\Y_{D,U}^{ij}$ are given by \cite{Badurina:2024nge} 
\begin{align}
    \!\sum \Y_U^{ij}  \!&= \! n(N+n-2N_L)(1-\cos(\vb{\qu} \cdot \vb{\Delta x})) ~,\label{eq:YUij}\\
    \hspace{-0.2cm} 
    \!\sum \Y_D^{ij}  \!&= \! n^2 (1-\cosqx)- \! \iu n N \! \sinqx ~,\quad \label{eq:YDij}\\
    \!\sum \Y_D^{ii} \!&= \!   |n| (1-\cosqx) - \iu n \sinqx~. \label{eq:YDii}
\end{align}

Next, let us consider $\rho^{(2)}_r$. We encounter new objects beyond $G_{ij}(\vb{q})$ because  correlation functions of the form,
\begin{equation}
  G_{ijk\ell}(\vb{q}_1,\vb{q}_2)= \left\langle  \e^{\iu \vb{q}_1 \cdot (\vb{y}_i-\vb{y}_j)} \e^{ \iu \vb{q}_2 \cdot (\vb{y}_k-\vb{y}_\ell)}\right\rangle_y ~
\end{equation}
appear, where the angle brackets denote averaging with respect to $\varrho_y(\{\vb{y}\},\{\vb{y}\})$.

For uncorrelated systems, such as the gas of non-interacting atoms we consider, one has that 
\begin{equation}
    G_{ijk\ell}(\vb{q}_1,\vb{q}_2) = G(\vb{q}_1) G(\vb{q}_2)\qq{for} i\neq j\neq k \neq \ell~. 
\end{equation}
When one or more indices are degenerate, however, this factorization is spoiled (even for uncorrelated states), which in turn spoils the exponentiation of $G(\vb{q})$ starting at $\rho_r^{(2)}$. By adding and subtracting the {\it missing} terms, required for exponentiation, we can write 
\begin{equation}\label{eq:exponentiation}
  \langle \e^{-\int_t \lambda} \rangle_y = \e^{-\int_t \langle \lambda \rangle_y} ( 1  + \delta^{(2)} + \delta^{(3)} +\cdots)~,
\end{equation}
where we have omitted $\delta^{(1)}$ since it vanishes by definition. The multiplicative corrections, $\delta^{(m)}$, are suppressed by $1/N^{m-1}$ with respect to the terms participating in the exponentiation due to combinatorics. 

In \cref{app:2order} we explicitly compute the leading correction, $\delta^{(2)}$, and find that it is non-vanishing in general. However, for $|n|=1$ measurements, $\delta^{(2)} = 0$  vanishes identically; we conjecture that this holds to all orders for $|n|=1$, but we do not attempt a proof. Furthermore, in the limit of a point-like cloud, $\varrho_y(\{\vb{y}\},\{\vb{y}\}) \rightarrow \prod_i \delta(\vb{y})$, the exponentiation is exact (as can be easily proven) and we obtain the point-like limit derived in Ref.~\cite{Badurina:2024nge}.

Let us now neglect $\delta^{(m)}$, such that we approximate \cref{eq:exponentiation} by its exponentiation form. Then the expectation value of the observable count-rate, $\langle O_+ \rangle_{t=\mathcal{T}}$, maps to \cref{best-fit-V-phi} with
\begin{equation}\label{eq:Vandphi}
    V = \e^{-(s+s_0)} \cos^{N-1}(\tau), \quad \text{ and }\quad \varphi = \gamma_0 + N \gamma~,
\end{equation}
with, 
\begin{align}
s_0 &= \int_{t,q} \omega_D(\vb{q}) (1-\cosqx) ( 1- \FF)~,\\
\gamma_0 &= \int_{t,q} \omega_D(\vb{q})\sinqx ( 1- \FF)~,\\
s &= \int_{t,q} \omega_D(\vb{q}) (1-\cosqx)  \FF~,\\
\gamma &= \int_{t,q} \omega_D(\vb{q})\sinqx  \FF~,\\
\tau &=  \int_{t,\qu} \omega_U(\vb{\qu})(1-\cos(\vb{\qu}\cdot\vb{\Delta x})) G(\vb{\qu})~.
\end{align}
Above, we have explicitly separated the incoherent ($s_0$ and $\gamma_0$) and coherent ($s$, $\gamma$ and $\tau$) contributions to the observables.
Notice that $s+s_0$ contains no coherent enhancements and reduces to the naive single-particle calculation \cite{Gallis:1990,Hornberger_2003}. For $n\geq2$ one must track $s$ and $s_0$ separately. If our conjecture that $\delta^{(m)}(|n|=1)=0$ for all $m$, then the above equations are exact for $\langle O_+ \rangle_{t=\mathcal{T}}$. Higher moments, such as the variance, will involve $|n|>1$-body density matrices where corrections in \cref{eq:exponentiation} will first appear at ${\cal O}(g^4 \mathcal{T}^2)$, i.e., $\delta^{(2)}\neq 0$. 

Let us briefly summarize the main results of this section. When the spreading of the cloud can be neglected the master equation, \cref{full-EOM}, can be solved by the method of characteristics giving \cref{eq:expl-soln-semiclassical}. When measurements are performed on the internal labels of the atom, we trace over $\{\vb{y}\}$ and obtain a reduced density matrix $\rho_r(\{s\},\{s'\})$. The trace over $\{\vb{y}\}$ introduces form factors, and higher order correlation functions. The resulting expression approximately exponentiates as discussed around \cref{eq:exponentiation}. 

We note that \cref{eq:lambdaG} maps to existing results in the literature in the appropriate kinematical regime. For example, in the point-like limit where the background does not resolve the interferometer's cloud, i.e. $q \ll 1/r_c$, the Debye-Waller factor in~\cref{eq:G} becomes the identity, and \cref{eq:YUij} and \cref{eq:YDij} are the only contributions to the reduced density matrix, which map to the unitary and decoherence kernels obtained in~\cite{Badurina:2024nge}. In this limit, only the coherent $s$, $\gamma$ and $\tau$ affect the visibility and phase-shift. Allowing the background to resolve the cloud, we have shown that, for 1-body measurements, the Debye-Waller factor exponentiates up to ${\cal O}(g^6 \mathcal{T}^3)$. This reproduces the formalism used in~\cite{Du:2022ceh,Du:2023eae} where, for $N \gg 1$, $G(\bf{q})$ appears as a form factor convoluting the differential interaction rate.

\section{Perturbative decoherence  \label{sec:naive-TDPT} }
%
Having solved the case of static cloud structure, we now consider approximations that are appropriate when including the spreading of the cloud as a function of time. 
For very feeble couplings between the atoms and the decohering background gas, we may study the hierarchy,
\begin{equation}
    \label{slow-hierarchy}
    t_{\rm kick} \sim  t_{\rm spread} \ll t_{\rm decoh.}~.
\end{equation}
In this limit, decoherence is weak and can be treated perturbatively on top of standard unitary evolution of the interferometer. 

Let us solve 
\begin{equation}
  \label{full-EOM-repeated}
  \pdv{\rho}{t} = -\iu \qty[ H_{\rm sys.}, \rho] - \qty(\lambda_U + \lambda_D)\rho~, 
\end{equation} 
perturbatively in $g$, counting $\lambda_U\sim \lambda_D \sim g^2$. We expand $\rho$ order-by-order in $g^2$
\begin{equation}
    \rho(t)= \rho^{(0)}(t) + \rho^{(1)}(t) + \ldots~.
\end{equation}

The zeroth order solution $\rho^{(0)}$ satisfies
\begin{equation}
    \partial_t \rho^{(0)}  + \iu [H_{\rm sys},\rho^{(0)}] = 0~,
\end{equation}
and is given explicitly by,
\begin{align}\label{eq:rho0}
    &\rho^{(0)}(\{\vb{x}\},\{\vb{x}'\},t) \\
    &= \!\! \int_{t, y,y'} \!\!\!\!\!\!\!\!\! G_0(\{\vb{x}-\vb{y}\},t)  \varrho_y(\{\vb{y}\},\{ \vb{y}'\} )G_0^*(\{\vb{x}'-\vb{y}'\},t)~,
    \nonumber
\end{align}
where $G_0$ is the free particle propagator, 
\begin{equation}
    \label{free-greens-function}
G_0(\{ \vb{x} - \vb{y}\} ,t) = \left(\frac{m}{2\pi \iu ~t}\right)^{\frac{3N}{2}} \prod_{i}^N 
\e^{\iu \frac{m(\vb{x}_i-\vb{y}_i)^2}{2t}}~.
\end{equation}
Next, $\rho^{(1)}$ is  obtained by solving, 
\begin{equation}
\partial_t \rho^{(1)} + \iu [H_{\rm sys},\rho^{(1)}] = -\lambda \rho^{(0)}~.
\end{equation}
The homogeneous solution vanishes because of boundary conditions of $\rho^{(1)}(t=0)=0$, and so we are left only with the particular solution,
\begin{widetext}
\begin{equation}
\begin{split}
\rho^{(1)} &(\{\vb{x}\},\{\vb{x'}\},t) = 
 -\int_0^t \dd t'\int_{ y, y'} \!\!\!  G_0(\{ \vb{x}-\vb{y}\}, t')  \lambda(\{\vb{y}\}, \{\vb{y}'\}, t') \rho^{(0)}(\{\vb{y}\},\{\vb{y}'\},t') G_0^*(\{\vb{x}'-\vb{y}'\},t') ~.
\end{split}
\end{equation}
Tracing over position, $\int_x$, at $t=\mathcal{T}$ gives, 
    \begin{equation}\label{eq:perturbative-decoherence-result}
    \begin{split}
    \rho_r(\{s\},\{s'\}) 
    &= \e^{\iu n \varphi_0}\varrho_s\qty(\{s\},\{s'\}) \qty[ 1 - \int_0^\mathcal{T} \dd t' \int_y~\lambda(\{\vb{x}_s(t')\},\{\vb{x}_{s'}(t')\}) \times \rho^{(0)}(\{\vb{y}\},\{\vb{y}\},t')] + O(g^4\mathcal{T}^2)~,
    \end{split}
    \end{equation}
\end{widetext}
where $\vb{x}_s(t)= \vb{Y}_s(t) + \vb{y}$ as given in \cref{comoving-coords}, and
 we have used the  identity 
\begin{equation}
\int_{x} G_0(\{\vb{x}-\vb{y}\}) G_0^*(\{\vb{x}-\vb{y'}\}) =  \prod_i^N \delta^{(3)} (\vb{y}_i - \vb{y}_i')~.
\end{equation}
\Cref{eq:perturbative-decoherence-result} can be re-written in the same form as \cref{eq:lambdaG}, but now with a time-dependent form factor, 

\begin{equation}\label{eq:Gtau}
\begin{split}
    G(\vb{q},t) 
    &= \int_y \rho^{(0)}(\{\vb{y}\},\{\vb{y}\},t)~ \e^{\iu \vb{q}\cdot(\vb{y}_i-\vb{y}_j)}~,~i\neq j~.
\end{split}
\end{equation}

Including higher orders in the $g^2\mathcal{T}$ expansion, the calculation becomes more involved. One can generically obtain the solution up to $\alpha$-th order by performing an integral of nested propagators and $\lambda$ factors that connect the initial condition $\varrho_y$ to $\rho_r$ as follows:
\begin{equation}
\begin{split}
&\rho_r \simeq 1 - \int_t \int_{y_1} \!\!\!\! \lambda(\{ \vb{y}_1\},\{ \vb{y}_1\}) \int_{y_2,y_2'} \!\!\!\!\!\! G_0(\{ \vb{y}_1 -\vb{y}_2)  \\
&\times  \bigg[  \varrho_y(\{\vb{y}_2\},\{\vb{y}_2'\}) - \lambda(\{ \vb{y}_2\},\{\vb{y}_2'\})   \int_{y_3,y_3'} \!\!\!\!\!\! G_0(\{ \vb{y}_2 -\vb{y}_3)  \\
& \times \bigg[ \varrho_y(\{ \vb{y}_3\}, \{ \vb{y}_3'\}) - \lambda(\{ \vb{y}_3\}, \{ \vb{y}_3'\}) \times \cdots \\
& \times  \bigg[ G_0(\{ \vb{y}_{\beta+1} - \vb{y}_{\beta}\}) \varrho_y(\{\vb{y}_\beta,\vb{y}'_\beta) G^*_0(\{\vb{y}'_{\beta + 1},\vb{y}'_\beta\})\bigg] \\
& \hspace{0.1\linewidth}\times  \cdots \bigg] G_0^*(\{\vb{y}_2' - \vb{y}_3'\}) \bigg] G_0^*(\{\vb{y}'_1-\vb{y}'_2\}) ~,
\end{split}
\end{equation}
where $\beta = 2 \alpha -1$, and we have dropped the time dependence of the functions for simplicity. At higher orders, effects cannot be captured with a time-dependent cloud form factor. 
We do not pursue this more general solution here, and content ourselves with the leading $O(g^2 \mathcal{T})$ contribution in \cref{eq:perturbative-decoherence-result} when considering decoherence rates that satisfy the hierarchy in \cref{slow-hierarchy}.

Again let us summarize our intermediate results. For weak couplings $g\ll 1$, decoherence can be approximated by \cref{eq:perturbative-decoherence-result} which incorporates the spreading of the atom cloud as a function of time. In their common domain of validity ($g\ll 1$, with negligible cloud spreading) \cref{eq:perturbative-decoherence-result,eq:expl-soln-semiclassical} agree with one another.  The results of \cref{sec:static} are best suited to scenarios in which decoherence is strong such that the full series of $g^2 \mathcal{T}$ effects must be resummed; the cost is that time-dependent cloud substructure has been neglected. The formulation in \cref{eq:perturbative-decoherence-result} is best suited when decoherence rates are small, or time dependent cloud substructure must be included; the cost is that one must work order-by-order in $g^2 \mathcal{T}$. To our knowledge, this is the first time that cloud spreading effects have been properly incorporated in the open-system formalism of collisional decoherence, and generalizes the results obtained in previous literature, e.g.~\cite{Riedel:2016acj,Du:2022ceh}. The mapping to the point-like limit studied in Ref.~\cite{Badurina:2024nge} is straightforward.

\section{Applications \label{sec:applications} }
Having derived formulae for the static approximation and using standard fixed-order perturbation theory, we now apply our results to some simple examples. First we consider a thermal cloud of atoms in a harmonic trap studying the time-dependent form factor in the context of time-dependent perturbation theory. Second, we consider a realistic pulse sequence (as used in current tabletop atom interferometers) and compute the coherently enhanced phase shift induced by a photon beam via Rayleigh scattering. We then perform a similar calculation for the phase shift induced by the scattering of electrons from neutral atoms (including both the coherent and incoherent parts). Next, we sketch how to apply our results to matter interferometers such as the proposed MAQRO experiment \cite{Kaltenbaek:2015kha,Kaltenbaek:2023xtz} or the microsphere proposal of Ref.~\cite{Pino_2018}. Finally, we outline how the results presented herein bear on the detection of dark matter interacting with atoms via long range forces. 

\subsection{Thermal cloud \label{sec:thermal-cloud}}
Consider a cloud of atoms initially prepared in a thermal state with temperature $\beta^{-1}$, using a harmonic oscillator trap $V(\vb{x}) = \tfrac12 M \omega^2 \vb{x}^2$ with an oscillator frequency $\omega$. The system of atoms is then described, at $t=0$, by a thermal density matrix, 
\begin{align}
    \label{thermal-state}
        &\varrho_y (\{ \vb{y} \},\{ \vb{y}'\}) =\prod_{i}^N \left[\frac{M \omega}{\pi}\tanh\left(\frac{\omega \beta}{2}\right)\right]^{\frac{3}{2}} \\
        &~\times\text{exp}\bigg[\!\!- \!\! \frac{M \omega/2}{\sinh (\omega \beta)} \bigg(  \!\! (\vb{y}_i^2 + \vb{y}_i'^2) \cosh (\omega \beta)- 2 \vb{y}_i \cdot \vb{y}_i'\!\bigg) \! \bigg]~.\nonumber
\end{align}
This then gives for the initial density distribution of the gas,
\begin{equation}
    \varrho_y(\{ \vb{y} \},\{ \vb{y}\}) = \prod_{i=1}^N \frac{\e^{- \frac12 \vb{y}_i^2/r_c^2}}{(\sqrt{2\pi}~r_c)^3}~,
\end{equation}
where the thermal cloud radius at $t=0$ is given by, 
\begin{equation}\label{eq:rc}
    r_c(t=0) =\frac{1}{ \sqrt{2M \, \omega \tanh (\beta \omega /2)}}~.
\end{equation}
Using \cref{thermal-state} and \cref{free-greens-function} we can obtain the time-dependence of the cloud form factor 
\begin{equation}\label{eq:tFFexplicit}
    G(\vb{q},t) = \e^{-\tfrac12 q^2 r_c^2(t) }~,  
\end{equation}
where $q = |\vb{q}|$. We follow this convention hereafter. The cloud radius evolves now as a function of time, 
\begin{equation}
    \begin{split}
        r_c(t) &= r_c(0) \times \sqrt{1+ \omega^2t^2}~.
    \end{split}
\end{equation}
Notice that the cloud's expansion rate is set by the oscillator frequency and not by the temperature \cite{Naraschewski:1996PRA}. The static approximation is valid when $\omega \mathcal{T} \ll 1$ and we may use $G(\vb{q},t=0)$.

All time-dependence in physical observables is contained in factors of $\sin(\vb{q}\cdot \vb*{\Delta}\vb{x})$, $\cos(\vb{q}\cdot \vb*{\Delta}\vb{x})$, and the time-dependent form factor $G(\vb{q},t)$, 
\begin{equation}\label{eq:Gqexp}
    G(\vb{q},t) = \exp[ -\tfrac12 q^2 r_0^2 (1+\omega^2t^2)] ~.
\end{equation}
Whenever $\vb*{\Delta}\vb{x}$ depends linearly on time (which is generic for atom interferometers), we can perform the time integral first and obtain an analytic result,
\begin{equation}
\begin{split}
\mathcal{I}(\vb{q}\cdot \vb{v})&= \int \dd t \, \e^{- \frac{q^2 \omega t^2}{4 M \tanh(\beta \omega/2)}} \e^{\iu \vb{q} \cdot \vb{v} t}\\
&= \iu \sqrt{\tfrac{\pi M\tanh(\beta \omega/2)}{q^2 \omega }}e^{ - \frac{M (\vb{q}\cdot \vb{v})^2 \tanh(\beta \omega /2)}{ q^2  \omega}}\\
&\quad \times \bigg( \text{erfi}\left[ \sqrt{\tfrac{M (\vb{q}\cdot \vb{v})^2 \tan(\beta \omega/2)}{q^2 \omega }}\right] \\
&\qquad - \text{erfi}\left[\tfrac{\iu q^2 \omega \mathcal{T} + 2 M  \vb{q}\cdot \vb{v}  \tanh(\beta \omega/2)}{2\sqrt{M q^2  \tanh(\beta \omega/2) \omega}} \right]\bigg)~.
\end{split}
\end{equation}
This result holds for arbitrary interactions (since we have performed the time integral first), and allows for the inclusion of cloud spreading in computations of both unitary and decohering effects. This may be relevant for e.g., the Stanford atom fountain, where the cloud expands by a factor of 30~\cite{Asenbaum:2020era,Overstreet:2022}. 

In practice, it is often easier to perform other integrations (e.g., over the momentum transfer) first. These integrals depend on the interaction between the background gas and the atoms. In what follows we discuss concrete physical examples for sake of illustration. 

\subsection{Phase shift from an IR lamp \label{sec:lammp-application}}

Let us illustrate the impact of a decohering probe on the interferometer observables defined above with an example. Consider a beam of photons traveling in the $\hat{\vb{z}}$ direction. We assume a spectrum $\dd \Phi/\dd E$ which when integrated gives a total flux. 

At long wavelengths, photons scatter from atoms via Rayleigh scattering, which may be classically understood as the scattering of light from an induced electric dipole. The Rayleigh‑scattering rate is given by:
\begin{equation}\label{eq:xsecRay}
    \frac{\dd \sigma}{\dd q^2} = \pi \alpha_{\rm pol}^2 E^2 \left(1+  \qty[1 - \frac{q^2}{2E^2}]^2\right)~.
\end{equation}
Here $E=k=|\vb{k}|$ is the energy of the incoming photon, and $\alpha_{\rm pol}$ is the polarizability of the atom. 

Since a phase shift is the cleanest observable to access experimentally, let us consider the coherently enhanced phase shift due to the scattering of a photon beam with the atoms, 
\begin{equation}
    \gamma = \!\!\! \int \!\! \dd t   \! \int  \!\! \dd E \, \frac{ \dd \Phi}{\dd E}\!  \int \! \dd q^2 \frac{\dd \sigma}{\dd q^2}G(\vb{q},t)\!\! \int \! \frac{\dd \phi}{2\pi} \sin(\vb{q} \! \cdot \vb{\Delta x}).
\end{equation}
In general, the pulse sequence is given by
\begin{equation}
    \vb*{\Delta} \vb{x}(t) = \vb{v} t~ \Theta(\mathcal{T}/2-t) + \vb{v}(\mathcal{T}- t) ~\Theta(t-\mathcal{T}/2)~.
\end{equation}
This describes an atom cloud which is kicked apart, then kicked again at $\mathcal{T}_{1/2} = \mathcal{T}/2$, and then recombined at the time of measurement $\mathcal{T}$. 
We note that the integral is symmetric in the two halves of the path when the spreading of the cloud is neglected. For the purpose of deriving analytical expressions for $\gamma$, in what follows we ignore the cloud spreading, i.e.,
\begin{equation}
    G(\vb{q},t) \rightarrow G(\vb{q}) = \e^{-\tfrac{1}{2}q^2 r_c^2}~,
\end{equation}
with $r_c$ given in \cref{eq:rc}.

The integral over $\phi$ can be readily performed, using $\gamma=2\times \gamma_{1/2}$ (as discussed above) and gives~\footnote{We use $k$ rather than $E$ in $\sin(q^2 v_z t/2 k)$, which is helpful when generalizing to non-relativistic electrons below.}
\begin{equation}\label{eq:gammaxsec}
\begin{split}
    \gamma = 2\times&\int_0^{\mathcal{T}_{1/2}}  \!\!\! \dd t \int \dd E \, \frac{\dd \Phi}{\dd E} \\
    & \!\!\! \times \int \dd q^2 \frac{\dd \sigma}{\dd q^2}  \, G(\vb{q}) \, J_0(q_\perp v_\perp t ) \, \sin \left (\frac{q^2 v_z t}{2k}\right)~.
\end{split}
\end{equation}
Above, 
\begin{equation}
    \begin{split}
    q_\perp &= q \sqrt{1-\frac{q^2}{4k^2}}~,
    \end{split}
\end{equation}
$J_0$ is a Bessel function of first kind, and $v_\perp^2 = v_x^2 + v_y^2$ and $v_z$ are related to the Cartesian components of $\vb{v}=(v_x,v_y,v_z)$.

For coherent scattering, with $q\ll 1/r_c\ll k$, we can use $q_\perp \simeq q$ and $\sin(q^2 v_z t /2k)\simeq q^2 v_z t /2k$. The latter holds for sufficiently short times i.e., one requires $q^2 v_z t /2k \ll 1$. This approximation is satisfied for $q \lesssim 1/r_c$ for the run-time benchmarks we choose. For the interaction shown in \cref{eq:xsecRay} we then find, 
\begin{equation}
\begin{split}
    \label{eq:small-q}
    \gamma &\simeq 2\pi \Phi \alpha_{\rm  pol}^2 v_z  \langle E \rangle \!\! \int \!\! \dd q \, q^3\e^{-\tfrac{1}{2}q^2 r_c^2} \!\! \int_0^{\mathcal{T}_{1/2}} \!\!\!\!\! \dd t\, t J_0(q v_\perp t)~,
\end{split}
\end{equation}
where $\langle E \rangle = (1/\Phi) \int \dd E \, E \, \dd \Phi/ \dd E$.
Since we begin with a thermal state of atoms prepared in a harmonic trap, as described in \cref{sec:thermal-cloud}, the form factor is given by \cref{eq:Gqexp}. We have also substituted the cross-section from \cref{eq:xsecRay} in the limit of $q \ll E$.  

Now we perform the time integral,
\begin{equation}\label{eq:timeintegral}
\begin{split}
    &\int_0^{\mathcal{T}_{1/2}} \dd t\,t \, J_0(q v_\perp t) = \frac{\mathcal{T}_{1/2} \, J_1(q \mathcal{T}_{1/2} v_\perp)}{q v_\perp} ~,
\end{split}
\end{equation}
where $J_1(x)$ is the derivative of $-J_0(x)$.
Finally, we integrate 
$q$ from 0 to $\infty$, since the cloud form factor itself provides a ultraviolet cutoff below the kinematic limit, obtaining
\begin{equation}\label{eq:gammaanaphot}
\begin{split}
    \gamma &\simeq 2 \pi \Phi \, \mathcal{T}_{1/2} \frac{\alpha_{\rm pol}^2}{r_c^4} \Delta x_z \langle E \rangle\e^{-\frac{\Delta x_\perp^2}{2 r_c^2}} ~,
\end{split}
\end{equation}
where $\Delta x_z = v_z\mathcal{T}_{1/2}$ and $\Delta x_\perp = v_\perp \mathcal{T}_{1/2}$ are the maximum separation of the two atom clouds in the $z$ direction and in its perpendicular plane, respectively. 

Before proceeding, let us discuss some interesting features of \cref{eq:gammaanaphot}. First, the $1/r_c^4$ prefactor can be easily understood in the $r_c \rightarrow \infty$ limit where only small-$q$ of order $1/r_c$ can contribute to the integral. The exponential suppression stems from the oscillatory behavior of the Bessel function $J_1(q \Delta x_\perp)$. It is worth emphasizing that this exponential dependence is directly related to the Gaussian form factor of a harmonically trapped gas. Therefore, one sees that the observable phase shift, $\gamma$, is sensitive to both the radius and shape of the atom cloud. 

Clearly, one requires $\Delta x_\perp \sim r_c$ or smaller to avoid exponential suppression of the phase shift.  \Cref{eq:gammaanaphot} already suggests the optimal experimental setup to measure the coherently enhanced phase at the laboratory. The penalty from the hierarchy of the scales $\Delta x_\perp$ and $r_c$ in the Bessel functions and the exponential can be mitigated by orienting the photon lamp relative to the interferometer such that \begin{equation}\label{eq:optimal_condition}
 \Delta x = \sqrt{\Delta x_z^2+ \Delta x_\perp^2} \sim |\Delta x_z| \gg \Delta x_\perp \sim r_c~,
\end{equation} 
where $\Delta x = |\vb{\Delta x}|$.

In the following we fix benchmark values for the parameters in $\gamma$ to estimate the magnitude of the phase shift. We imagine an array of LED infrared lamps of $\sim 10~{\rm mm} \times 10~{\rm mm}$. Motivated by specifications from Ref.~\cite{IRlamp} (with each mm$^2$ lamp costing $\sim 1$ USD) we estimate a power per unit area of $P_{\rm LED} = 4\pi \times (500~{\rm mW}/{\rm mm}^2)$ and a photon wavelength $\lambda = 850$ nm. This corresponds to $\langle E \rangle = 2\pi/\lambda = 7392~{\rm mm}^{-1}$ (equivalent to 1.459 eV) and a flux of $\Phi\sim 1.3 \times 10^{19}~{\rm mm}^{-2}{\rm s}^{-1}$. We take $\Delta x_z  \sim 1 \text{ cm}$, $\Delta x_\perp \sim 0.1 \text{ mm}$ , $r_c \sim 100 \, \mu\text{m}$, and $\mathcal{T}_{1/2} \sim 0.1\text{s}$.\!\footnote{These benchmarks correspond to recoil velocities of $v_z \sim 10 \text{ cm/s}$ and $v_\perp \sim 1 \text{ mm/s}$.} 

We assume $N \sim 10^6$ atoms of $\phantom{}^{87}\text{Rb}$ with polarizability $\alpha = 47.39(8) ~\text{\AA}^3$~\cite{Gregoire_2015}. Plugging these benchmarks into \cref{eq:gammaanaphot} we obtain, 
\begin{align}
    \varphi  \simeq \gamma N &\sim 8.2 \times 10^{-6} \left(\frac{P_{\rm LED}}{6 ~{\rm W}/{\rm mm}^{2}}\right) \left(\frac{\mathcal{T}_{1/2}}{0.1 \text{ s}}\right)\\ 
    & \times \left(\frac{\Delta x_z}{1 \text{ cm}}\right) \left(\frac{850 \text{ nm}}{\lambda}\right) \left(\frac{\alpha_{\rm pol}}{47.39 ~\text{\AA}^3}\right)^2 \left(\frac{N}{10^6}\right) ~,\nonumber
\end{align}
which lies below shot noise $\delta \varphi_{\rm shot} =10^{-3} \,\sqrt{10^6/N}$ for a million atoms for a single experimental run. Repeating the experiment $\aleph$ times would reduce the error as $1/\sqrt{\aleph}$ until systematics take over, i.e.,
\begin{equation}\label{eq:statistics}
    \delta \varphi (\aleph) = \sqrt{\frac{\delta \varphi_{\rm shot}^2}{\aleph} + \delta \varphi_{\rm sys}^2}~.
\end{equation}
The effect should therefore be observable for a live run-time in excess of 1 hour. We do not consider a more detailed optimization of the lamp's wavelength (and  potential challenges such as inelastic excitations of the atom cloud).   We note that the coherent effect is {\it much} larger than that from incoherent Rayleigh scattering. We return to incoherent scattering below in the context of an electron gun where the incoherent contribution is much larger.

\subsection{Phase shift from an electron beam \label{sec:e-application}}
\begin{figure}[t]
\includegraphics[width=\linewidth]{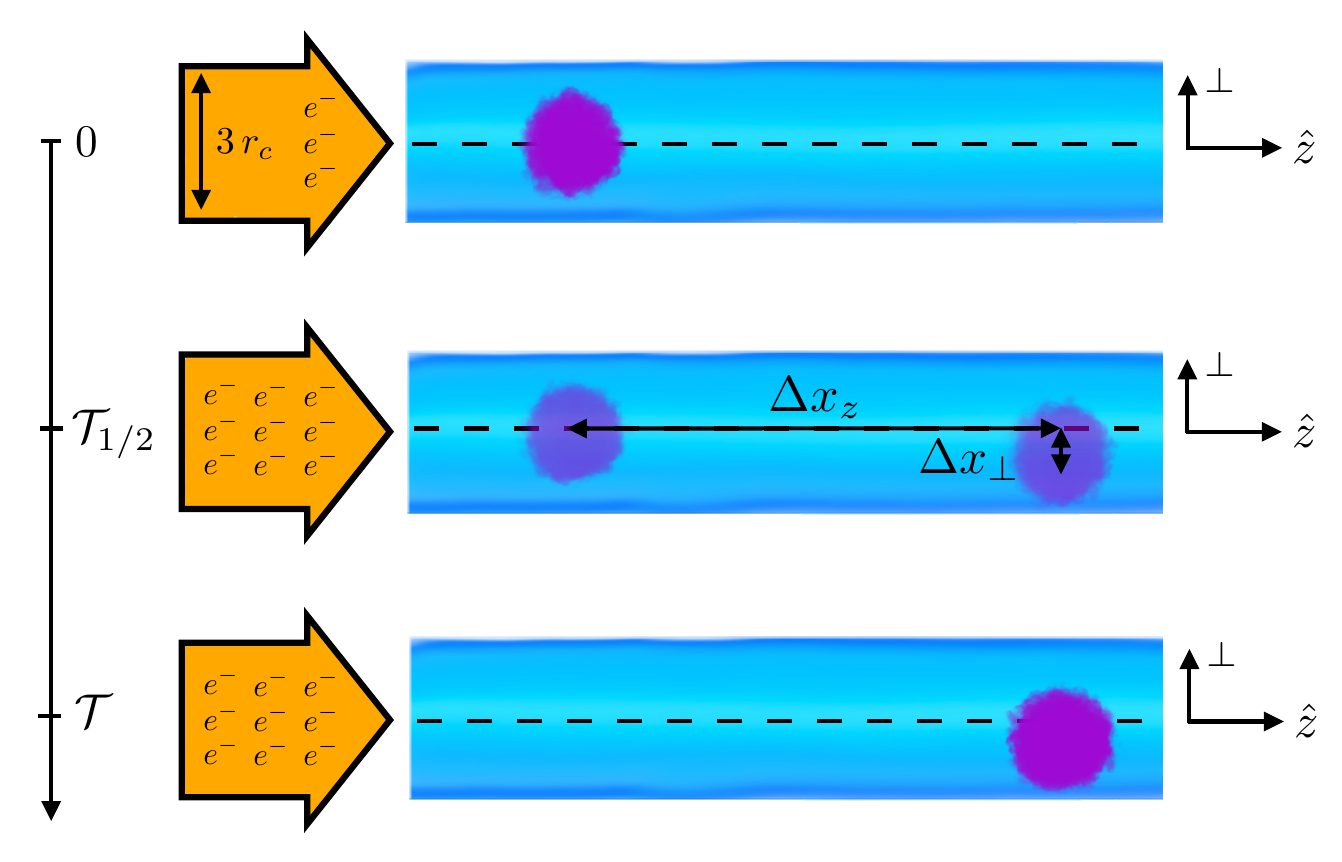}
\caption{A schematic of the electron gun setup we consider (the infrared lamp is conceptually similar). A $1~{\rm eV}$ energy electron gun produces a beam of radius $\sim 3 r_c$. The beam is directed nearly parallel to the path of an atom interferometer. During the pulse sequence, each atom is placed in a superposition with one branch, $L$, moving forward between $0$ and $\mathcal{T}_{1/2}$ while the other, $R$, is static. A $\pi$ pulse is applied at $\mathcal{T}_{1/2}$ (with the atom clouds in quantum superposition)  after which the $L$-branch comes to rest, while the $R$-branch is kicked. At time $\mathcal{T}$, both branches are coincidentally located at $\vb*{\Delta x}$ and a measurement is performed with the $\ket{\pm}$ states. \label{fig:electrongun} }
\end{figure}

Next, let us consider an alternative to the infrared lamp discussed above. In particular, consider an electron gun/beam~\cite{kimball_electron_gun_systems} with low-energy ($\sim 1~{\rm eV}$ kinetic energy) electrons which scatter off of the neutral atoms in the interferometer as shown in \cref{fig:electrongun}. 

Electrons interact with atoms by photon exchange. There are two dominant scattering mechanisms: {\it i)} the monopole scattering off the neutral atom, and {\it ii)} dipole scattering off the atom's magnetic dipole moment. In what follows, we consider the former interaction because we are interested in interferometers that rely on hyperfine splittings of states with total angular momentum 
$F=0$; in such states the magnetic dipole moment of the atom vanishes.\!\footnote{Dipole scattering from an interferometer with $m_F\neq0$ states is another platform that could test the results of this paper.}

In the Born approximation, the screening of the nuclear charge by the atomic electrons is parametrized by an atomic form factor, which at low $q$ scales as
\begin{equation}
    F_A(\vb{q}) = q^2 r_A^2 + O(q^4)~,
\end{equation}
where $r_A$ has dimensions of length. For our numerical estimates, we extract $r_A$ from the total $e^-~^{87}{\rm Rb}$ cross section plotted in Fig.~7 of Ref.~\cite{PhysRevA.44.5693}, which goes down to $15~{\rm eV}$. Using an estimate of $\sigma \simeq 100~a_0^2$, we obtain\footnote{We set equal the integral of \cref{eq:xsec} from $q=0$ to $q=2p_e$  $\sigma=16\pi Z^2 r_A^4 \alpha_{\rm EM}^2 m_e^2$ and the estimate taken from Ref.~\cite{PhysRevA.44.5693} to arrive at this number.} $r_A \simeq (1/20~{\rm keV})$; we note that this is numerically similar to $a_0/\sqrt{Z} = 1/(22.7~{\rm keV})$.

Although the Born approximation breaks down at low velocities, the above discussion motivates us to model the shape of $\dd \sigma/\dd q^2$ by a contact interaction (i.e., flat in $q^2$). With this as a model, the differential cross section for elastic monopole scattering on an atom is then given by
\begin{equation}\label{eq:xsec}
    \frac{\dd \sigma}{\dd q^2} = |F_A(\vb{q})|^2 \left. \frac{\dd\sigma}{\dd q^2}\right|_{\rm Ruth.}\!\!\! \sim 4\pi Z^2 r_A^4 \alpha_{\rm EM}^2 \frac{1}{\beta_e^2}~,
\end{equation}
where we have used the Rutherford scattering with the fine-structure constant $\alpha_{\rm EM}$, 
\begin{equation}
   \left. \frac{\dd \sigma}{\dd q^2}\right|_{\rm Ruth.} \!\!= ~\frac{4\pi  Z^2 \alpha_{\rm EM}^2}{q^4} \frac{1}{\beta_e^2}~.
\end{equation}
Here, $\beta_e = p_e/m_e$, where $p_e$ is the momentum of the incoming electron and $m_e$, its mass. 
In the above equations we are assuming that the electrons do not resolve the nucleus. We note that, at $q \ll 1/r_A$, the differential cross-section behaves as a contact interaction.

Starting from \cref{eq:gammaxsec}, using $p_e$ in place of $k$, and the small-$q$ approximations discussed above \cref{eq:small-q} (specifically $q\ll p_e$), we find 
\begin{equation}
\begin{split}
    \gamma &\simeq 8\pi \, \Phi \, \alpha_{\rm EM}^2 Z^2 \frac{v_z r_A^4}{\beta_e^2 \, p_e} \int  \dd t \, t \\
    & \hspace{0.2\linewidth} \times  \int \dd q \, q^3 J_0(q v_\perp t)\e^{-\tfrac{1}{2}q^2 r_c^2}~,
\end{split}
\end{equation}
where we assumed, for simplicity, a sharp differential flux $\dd \Phi / \dd E = \Phi \, \delta(E - E_e)$. We note that a generic spectrum can be trivially incorporated by averaging $\langle \gamma(E_e) \rangle =(1/\Phi)\int \dd E \, \gamma(E_e) \dd \Phi / \dd E $.

Using the same steps as in the case of Rayleigh scattering but with the interaction given in \cref{eq:xsec}, we obtain 
\begin{equation}\label{eq:gammaana}
\begin{split}
    \gamma &= 8 \pi \Phi \, \alpha_{\rm EM}^2 Z^2 \! \left(\frac{r_A}{r_c}\right)^4  \frac{\mathcal{T}_{1/2} \Delta x_z}{\beta_e^2 \, p_e}  \, \text{exp}\left[- \frac{\Delta x_\perp^2}{2 r_c^2}\right] ~.
\end{split}
\end{equation}

Since electrons have momentum $p_e = \sqrt{2 m_e E} \sim ~{\rm keV} \,(\tfrac{E}{1~{\rm eV}})^{1/2}$ as compared to photons with $k \sim E$, momentum transfers can be much larger in electron scattering, which turns out to enhance the  phase shift induced by incoherent scattering. 

Consider the incoherent phase shift, $\gamma_0$, which can be conveniently expressed in terms of the quantity $\tilde{\gamma}$ defined by, 
\begin{equation}
    \label{incoh-e-gun}
    \tilde \gamma= 2\times \! \! \int_0^{\mathcal{T}_{1/2}} \!\!\!\! \dd t \! \int \dd E \frac{\dd \Phi}{\dd E} \!\! \int dq^2 \frac{\dd \sigma}{\dd q^2} \! \int \frac{\dd \phi}{2\pi}\sin(\vb{q}\cdot \vb{v} t)~.
\end{equation}
For momentum transfers larger than some threshold $q_{\rm loss}$, the atoms are kicked into a different velocity class and lost from the experiment \cite{letokhov2007laser,Cronin:2009zz}. The concrete value of $q_{\rm loss}$ depends on the detuning of the laser, the photon recoil momentum and the atom species employed. We take as a benchmark $q_{\rm loss}\sim 10~{\rm eV}$ and integrate up to this value rather than the kinematic limit of $q_{\rm max} = 2 p_e$.

The incoherent phase shift $\gamma_0$ is obtained from this expression, and the coherent phase shift $\gamma$, via, 
\begin{equation}
    \gamma_0 = \tilde{\gamma}-\gamma~.
\end{equation}
Since $\tilde{\gamma}$ is easier to compute, and since $\gamma$ is computed above, we focus on $\tilde{\gamma}$.

The integral in \cref{incoh-e-gun} accesses large momentum transfer for which $\sin(q^2 v_z t/q_{\rm max})$ cannot be Taylor expanded. Instead it is dominated by large $q\sim q_{\rm max}$ where rapid oscillations onset and a large-$\mathcal{T}_{1/2}$ analysis is appropriate. The analysis is best done with the $ \int \dd \phi \int \dd t$ integrals being performed first.  In the large $\mathcal{T}_{1/2}$ limit, we may use, for $a>0$ (see \cref{app:tricky-integral})
\begin{equation}\label{tricky-integral}
    \int \frac{\dd \phi}{(2\pi)} \int_0^{\mathcal{T}_{1/2}}
    \!\!\!\! \dd t \sin(a t + b t \cos\phi ) \rightarrow \frac{\Theta(a-b)}{\sqrt{a^2-b^2}} ~, 
\end{equation}
where $a=q_z v_z = v_z q^2/q_{\rm max}$ and $b=q_\perp v_\perp = qv_\perp\sqrt{1-q^2/q_{\rm max}^2}$ with $q_{\rm max} = 2p_e$. This formula is valid for $\mathcal{T}_{1/2}\rightarrow \infty$ with $a$, $b$, and $a-b\neq 0$ held fixed. 

Using, 
\begin{equation}
    \sqrt{a^2-b^2}= \frac{q^2}{q_{\rm max}} \sqrt{v_\perp^2 + v_z^2 - \frac{q^2_{\rm max}}{q^2} v_\perp^2}~,
\end{equation}
and that $a > b$, as enforced by $\Theta(a-b)$, implies that $q > q_{\rm max} v_\perp/\sqrt{v_\perp^2 + v_z^2}$, and we obtain
\begin{equation}
\tilde \gamma \simeq 32 \pi \Phi Z^2 \alpha_{\rm EM}^2 \frac{r_A^4}{v} \frac{p_e}{\beta_e^2} \! \int_{q_{\rm max}\frac{v_\perp}{v}}^{q_{\rm loss}} \!\! \frac{\dd q}{q}\frac{1}{\sqrt{1 - \frac{q_{\rm max}^2}{q^2} \frac{v_\perp^2}{v^2}}}~,
\end{equation}
where $v = |\vb{v}|$. The $q$ integral gives
\begin{equation}
\int_{q_{\rm max} \frac{v_\perp}{|\vb{v}|}}^{q_{\rm loss}} \frac{\dd q}{q} \frac{1}{\sqrt{1 - \frac{q_{\rm max}^2}{q^2} \frac{v_\perp^2}{v^2}}} ={\rm arccosh}\qty(\frac{v~q_{\rm loss}}{v_\perp q_{\rm max}})~.
\end{equation}
All together, $\tilde \gamma$ reads
\begin{equation}
\begin{split}
\tilde \gamma &\simeq 32 \pi \alpha_{\rm EM}^2 \Phi Z^2   r_A^4 \frac{p_e}{\beta_e^2} \,{\rm arcosh} \left(\frac{\Delta x~ q_{\rm loss} }{\Delta x_\perp q_{\rm max}}\right) \frac{\mathcal{T}_{1/2}}{\Delta x}~,
\end{split}
\end{equation}
and therefore
\begin{equation}
    \begin{split}
    \hspace{-0.04\linewidth} \gamma_0 = 8\pi \Phi &\frac{Z^2\alpha_{\rm EM}^2 }{\beta_e^2} r_A^2 \mathcal{T}_{1/2} \bigg\{\frac{4r_A^2 p_e}{\Delta x}{\rm arcosh} \left[\frac{\Delta x~ q_{\rm loss} }{\Delta x_\perp q_{\rm max}}\right] \\
    &\hspace{0.275\linewidth}- \frac{ r_A^2 \Delta x_z}{p_e r_c^4} \exp\qty[ -\frac{\Delta x_\perp^2}{r_c^2}] \bigg\}~.
    \end{split}
\end{equation}
We may now compare the contribution from $\tilde{\gamma}$ and $\gamma$ (the high-$q$ and low-$q$ contributions, respectively). The relevant ratio is~\footnote{Notice that for Rayleigh scattering $p_e\rightarrow k$ and since $k \sim 1~{\rm eV}$ the ratio $\tilde{\gamma}/\gamma_0$ is small.}
\begin{equation}
   \frac{\tilde{\gamma}}{\gamma} \sim (p_e r_c)^2 \times \qty(\frac{ r_c}{\Delta x})^2~, 
\end{equation}
where, following \cref{eq:optimal_condition}, we have assumed $\Delta x_z \sim \Delta x$. For our benchmarks, $p_e r_c \sim 10^6$, $r_c/\Delta x \sim 10^{-4}$, and $N\gtrsim 10^{5}$, we find $N\gamma \gtrsim \gamma_0$, i.e. the coherent contribution is larger than the incoherent contribution. 

Let us illustrate this explicitly with benchmark atom interferometer inputs. We use slightly different benchmarks as compared to the photon lamp, that serve to allow for both coherent and incoherent effects to compete for $N\sim 10^{6}$ atoms. We assume a cloud of $\phantom{}^{87}\text{Rb}$ with a radius of $r_c \sim 100 \, \mu\text{m}$, $\Delta x_z=1~{\rm m}$ and $\Delta x_\perp=0.1~{\rm mm}$, and a time of flight of $\mathcal{T}_{1/2} \sim 0.1 \text{ s}$. We consider an electron gun with a sharply peaked energy spectrum around $E \sim 1~\text{eV}$ and  a current of $I_e \sim 1~\text{A}$, see e.g., Ref.~\cite{kimball_electron_gun_systems} for a commercially available option. We imagine a configuration (shown explicitly in \cref{fig:electrongun}) where the beam width is set to $\sim 10 \pi r_c^2$ so as to fully cover the atom cloud, even at $\mathcal{T}_{1/2}$ where $x_\perp \sim 0.1~{\rm mm}$. This gives 
\begin{align}
    \Phi= \frac{I}{Q_e} \times \frac{1}{10\pi r_c^2}~,
\end{align}
where $I$ is the current and $Q_e$ is the charge of an electron.

Plugging numbers into our analytical expressions for $N\gamma$, and setting $\Delta x_\perp = r_c$, we get 
\begin{equation}
\begin{split}
  \hspace{-0.4cm} N \gamma  \simeq 6.4 \! \times \! 10^{-3} \!\left( \! \frac{\Delta x_z}{\, {\rm m}} \! \right)  \!\! \left(\! \frac{\mathcal{T}_{1/2}}{0.1 \, {\rm s}} \! \right)\!\! \left(\! \frac{N}{ 10^6} \! \right) \!\! \left(\! \frac{I_e}{{\rm A}}\! \right) \!\! \left( \! \frac{0.1\, \text{mm}}{r_c} \! \right)^6 \!\! \!,
\end{split}
\end{equation}
while for the incoherent phase shift we find, 
\begin{equation}
    \gamma_0 \simeq 5.1 \times 10^{-4} \left(\! \frac{\mathcal{T}_{1/2}}{0.1 \text{ s}} \! \right) \! \left( \frac{\text{m}}{\Delta x}  \right) \!\! \left( \! \frac{I_e}{\text{A}} \! \right) \!\! \left(\! \frac{0.1 \,\text{mm}}{r_c} \! \right)^2 \! \! .
\end{equation}
For an ampere current beam we expect a measurable effect from both the incoherent and coherent phase shifts within a single shot, where we have in mind a benchmark sensitivity of $1/\sqrt{N} = 10^{-3}$ per shot. By tuning the number of atoms in the cloud, the $N$-dependence of the coherent contribution can be tested.

\subsection{Matter interferometers}
Although we have developed the above formalism with atom interferometers in mind, the results of \cref{sec:static} apply immediately to matter interferometers. A matter interferometer involves  a nanosphere, or other macroscopic object, whose center of mass is placed in a quantum superposition. For simplicity, consider a pure state $\ket{\sf \Psi}$ with zero center of mass momentum,  which is then mapped to 
\begin{equation}
    \ket{\sf \Psi}\rightarrow \frac{1}{\sqrt{2}}\qty(\e^{\iu \vb{k} \cdot \hat{\vb{X} }}\ket{\sf \Psi} +  \e^{-\iu \vb{k} \cdot \hat{\vb{X} }}\ket{\sf \Psi})~. 
\end{equation}
The center of mass coordinate operator $\hat{\vb{X}}$ generates translations in momentum space. Although the Hamiltonian and many-body wavefunction that describes e.g., a gold nanosphere is complex when expressed in terms of its relative coordinates, the center of mass Hamiltonian is simply given by $\vb{P}^2/2M$. 

\Cref{eq:expl-soln-semiclassical} can be applied directly to solve for $\rho(\mathcal{T})$. When the Born approximation applies (e.g., in the case of dark matter scattering) \cref{eq:lambdaU,eq:lambdaD} can be used. If one considers scattering of other more strongly interacting particles such as neutrons, then \cref{eq:expl-soln-semiclassical} still applies, but the full induced $T$-matrix, $\mathcal{T}_{\{\vb{x}\}}$ should be used.\!\footnote{The full $T$-matrix can be obtained by treating the nanosphere as a background potential, solving the Schr\"odinger equation numerically for neutron scattering states, and extracting the relevant phase shifts for each partial wave.}

\subsection{Long-range dark matter scattering}
Atom interferometers have the potential to supply world-leading sensitivity to certain models of dark matter. In particular, if dark matter interacts with ordinary matter via a light mediator with feeble couplings (i.e., a long-range but very weak force), then its effective cross section in a conventional direct detection experiment, with an energy threshold $E_{\rm thr} \gtrsim 1~{\rm keV}$, can be much smaller than for a threshold-less atom interferometer. This naturally motivates the study of Yukawa-like interactions at low momentum transfers where coherent enhancements naturally occur. 

Consider a dark matter particle, $\chi$, that interacts with ordinary matter via a scalar $\phi$ with a mass $m_\phi$. For dark matter to be ``particle like'' one requires $m_\chi\gtrsim 10~{\rm eV}$ (below which dark matter begins to be wave-like)~\cite{Cheong:2024ose}. For typical virial velocities of $\sim 10^{-3}$ this implies that $p_\chi \gtrsim 10~{\rm meV}$.  The exchange of $\phi$ between a complex scalar, $\chi$, and ordinary matter leads to a Yukawa potential in the non-relativistic limit. For concreteness we consider a coupling to nucleons \cite{Knapen:2017xzo} 
\begin{equation}
   - \mathcal{L} \supset  2 m_\chi g_\chi \chi^\dagger \chi \phi  +  g_B  (\bar{p} p  + \bar{n} n) \phi ~, 
\end{equation}
where $g_B$ is the coupling of $\phi$ to baryons, and $g_\chi$ the coupling to dark matter. The nucleon-dark matter differential cross section is then given, for $m_\chi\ll (A m_n)$, by 
\begin{equation}\label{eq:DMxsec}
    \dv{\sigma}{q^2} = 4\pi A^2  \, \alpha_{\rm DM}^2 \frac{1}{u_\chi^2} \qty(\frac{1}{q^2+m_\phi^2})^2~,
\end{equation}
with $u_\chi$ the dark matter speed, $\alpha_{\rm DM} = g_\chi g_B / 4\pi$, and $A$ the atomic number, e.g., $A = 87$ for $\phantom{}^{87}\text{Rb}$.

Consider a ``cell'' of dark matter with number density $n_\chi$ traveling with velocity $\vb{u}_\chi$ towards the atom interferometer, and therefore a flux of $\chi$ given by $\Phi= n_\chi u_\chi$. Taking the particles of interest to be a subcomponent, $f_\chi \leq 1$, of dark matter we have $n_\chi = f_\chi \rho_\chi/m_\chi$, with $\rho_\chi =0.3~{\rm GeV}/{\rm cm}^3$. Therefore 
\begin{align}\label{eq:gammachi}
   \hspace{-0.05cm} \gamma({\vb{u}_\chi})&= 2 \times \frac{f_\chi \rho_\chi}{m_\chi^2} (4\pi) A^2 \alpha_{\rm DM}^2 \frac{\vb{u}_\chi \cdot \vb{v}}{u_\chi^3}
   \\
    & \!\!\!\!\!  \times \int_0^\infty \dd q \frac{q^3}{(q^2 + m_\phi^2)^2}\e^{-\tfrac{1}{2} q^2 r_c^2 }\int_0^{\mathcal{T}_{1/2}} \dd t  \, t \, J_0(q v_\perp t) ~, \nonumber
\end{align}
where $v_\perp = |\vb{v} - \hat{\vb{u}}_\chi \cdot \vb{v}|$ with $\hat{\vb{u}}_\chi$ a unit vector. Since $p_\chi \gtrsim 10~{\rm meV}$,  and we consider $r_c\sim 100~\mu{\rm m}$, it is safe to assume that $q \ll p_\chi = m_\chi u_\chi$ in the coherent regime (since $q r_c \lesssim 1$ as enforced by the cloud form factor and low dark matter velocities do not contribute substantially to the velocity average). As above, we ignore cloud spreading.

For $r_c m_\phi < 1$, we arrive at (see \cref{app:LRDM-integral}) 
\begin{equation}
    \label{phase-shift-LRDM}
    \hspace{-0.02\linewidth}\gamma(\vb{u}_\chi) \simeq 2\pi A^2 \alpha_{\rm DM}^2  \frac{f_\chi \rho_\chi}{m_\chi^2} \mathcal{T}_{1/2}\frac{\vb{u}_\chi\cdot\vb*{\Delta x}}{u_\chi^3} \log \bigg(\frac{1}{m_\phi^2 r_c^2}\bigg).
\end{equation}
For $r_c m_\phi \gg 1$, the results resemble the electron gun and Rayleigh scattering calculation and we instead obtain,  
\begin{equation}    \label{phase-shift-HMDM}
\begin{split}
    \gamma(\vb{u}_\chi)&\simeq 8 \pi \frac{f_\chi \rho_\chi}{m_\chi^2} A^2 \alpha_{\rm DM}^2 \left(\frac{1}{m_\phi r_c}\right)^4 \mathcal{T}_{1/2} \,     \\
    &\hspace{2cm} \times \frac{\vb{u}_\chi \cdot \vb{\Delta x}}{u_\chi^3}\e^{-\frac{1}{2} (1-[\vb{\hat u}_\chi \cdot \vb{\hat{v}}]^2)} ~,
\end{split}
\end{equation}
where we have used that $\Delta x_\perp \sim r_c \ll \Delta x_z$.

\Cref{phase-shift-LRDM,phase-shift-HMDM} are only valid for 
\begin{equation}\label{eq:argsin}
    v_\perp \mathcal{T}_{1/2} / (q_{\rm max} r_c^2) \ll 1~.
\end{equation}
Otherwise, the integration domain spans regions where the integrand becomes highly oscillatory, leading to cancellations and a suppression of the phase shift. For \cref{phase-shift-LRDM} the suppression is mild, and amounts to a slightly different argument of the logarithm, whereas for \cref{phase-shift-HMDM} the suppression is sizeable. For very small dark matter velocities, $u_\chi$, \cref{eq:argsin} can be violated (where $q_{\rm max} = 2 m_\chi u_\chi$), but these come from a negligibly small phase space volume and can be neglected.  

We must average over the velocity distribution for $\vb{u}_\chi$. For this purpose, we require the local dark matter velocity distribution, which is typically treated in the ``standard Halo model'' (see \cite{Baxter:2021pqo} and references therein)
\begin{equation}
    \dd n_\chi =  \frac{n_\chi}{\mathcal{N}} \e^{- U_\chi^2/u_0^2} \, \Theta(U_{\rm esc} -U_\chi)\, \dd^3 U_\chi~,
\end{equation}
where $U_{\rm esc}\approx 550~{\rm km}/{\rm s}$ is the magnitude of the galactic escape velocity, $\vb{u}_0$ is the local-standard-of-rest velocity with magnitude $u_0=|\vb{u}_0|\approx 240~{\rm km/s}$. The normalization constant, $\mathcal{N}$, is  defined such that $\int \dd^3 u_\chi  \frac{\dd n_\chi}{\dd^3 U_\chi} = n_\chi$. The velocity $\vb{U}_\chi$ is defined in the galactic frame and is related to the lab-frame velocity $\vb{u}_\chi$ by 
\begin{equation}
    \vb{U}_\chi = \vb{u}_\chi + \vb{u}_0~.
\end{equation}
We have neglected the Earth's velocity relative to the Sun and the Sun's peculiar velocity (small $\sim 10\%$ corrections as compared to the local-standard-of-rest velocity $\vb{u}_0$). 

For our purposes,  in the light mediator mass regime, $m_\phi r_c < 1$, we require the 
\begin{equation}
    \left\langle \frac{\vb{u}_\chi\cdot\vb*{\Delta x}}{u_\chi^3} \right\rangle = \frac{1}{\mathcal{N}} \int_{u_\chi<U_{\rm esc.}} \hspace{-0.1\linewidth}\dd^3 u_\chi \qty(\frac{\vb{u}_\chi\cdot\vb*{\Delta x}}{u_\chi^3} )\e^{- U_\chi^2/u_0^2} ~.
\end{equation}

Evaluating this average numerically we find 
\begin{equation}\label{eq:average_lightmediator}
     \left\langle \frac{\vb{u}_\chi\cdot\vb*{\Delta x}}{u_\chi^3} \right\rangle  
     =(- 0.43) \times \frac{\vb{u}_0\cdot\vb*{\Delta x}}{u_0^3}~.
\end{equation}
Similarly, for the large mediator mass regime, $m_\chi r_c \gg 1$, we require a more complicated average which is a function of $\vb{u}_0 \cdot \vb*{\Delta x}$. Sensitivity is maximized when $\vb*{\Delta x} \propto \vb{u}_0$ for which one has 
\begin{equation}
    \left \langle \frac{\vb{u}_\chi \cdot \vb{\Delta x}}{u_\chi^3}\e^{-\frac{1}{2} (1-[\vb{\hat u}_\chi \cdot \vb{\hat{v}}]^2)}\right \rangle_{\rm aligned} = (\pm 0.37) \times \frac{\Delta x}{u_0^2}~,
\end{equation}
with the relative sign determined by the orientation of the interferometer relative to the dark matter wind.

\begin{figure}[t]
    \centering
    \includegraphics[width=\columnwidth]{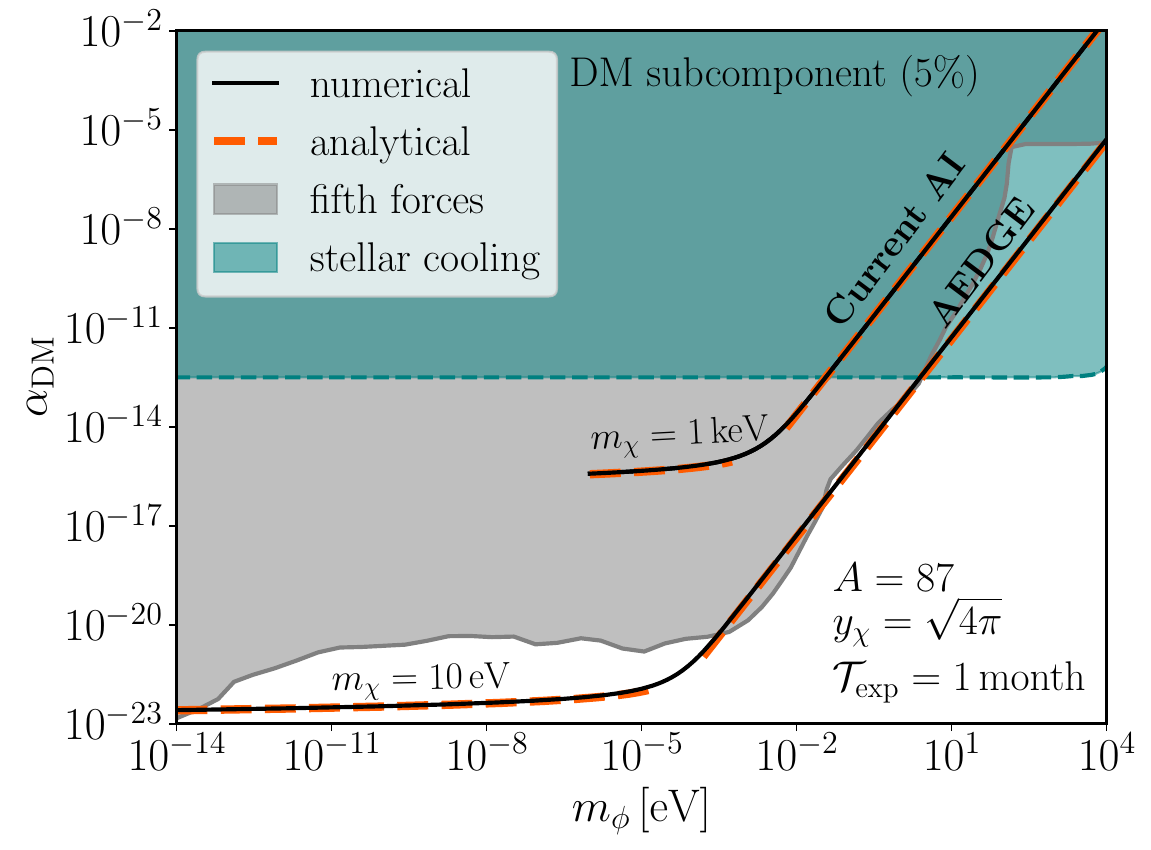}
    \caption{ Sensitivity projections for anomalous phase shift searches with atom interferometers. The curve labeled “Current AI” corresponds to typical benchmarks for table-top atom interferometers currently taking data (e.g., at Laboratoire Kastler Brossel~\cite{Morel:2020dww}, Berkeley~\cite{Parker:2018vye}, and the Stanford atomic fountain~\cite{Asenbaum:2020era}, among others): $m_\chi = 1 \text{ keV}$, $r_c = 100 ~\mu \text{m}$, $\mathcal{T}_{1/2} = 0.1 ~{\rm s}$, $\Delta x = 1 \text{ cm}$, $N=10^6$. 
    The curve labeled “AEDGE” follows the benchmarks quoted for the proposed space mission AEDGE~\cite{AEDGE:2019nxb}, with $N=10^{10}$, $r_c= 4 \, \text{mm}$, $\mathcal{T}_{1/2} = 600\text{ s}$, $\Delta x = 0.9 \text{ m}$. The red dashed curves show our analytical results while the black curves are obtained numerically.
   For all curves, we take $y_\chi = \sqrt{4\pi}$, $\Delta x_\perp = r_c$, an atomic species with $A=87$ nucleons (e.g. $\phantom{}^{87}\text{Rb}$ or $\phantom{}^{87}\text{Sr}$), and we assume that the experiment collects 1 month of data. 
    The gray shaded region shows existing constraints from fifth-force searches~\cite{Murata:2014nra}, while the teal shaded region shows constraints from anomalous stellar cooling~\cite{Hardy:2016kme}.
We assume the dark matter is a subcomponent with $\rho_{\chi} = 0.05 ~\rho_{\rm DM}$, in order to evade the strong self-interaction dark matter constraints at low dark matter masses \cite{Tulin:2017ara}.}
    \label{fig:plot}
\end{figure}

Assuming that the experiment collects data for $\mathcal{T}_{\rm exp}$ (repeating the experiment $\aleph = \mathcal{T}_{\rm exp}/\mathcal{T}$ times) and is limited by shot noise (i.e., prior to reaching a systematic floor), we apply \cref{eq:statistics} to derive the bounds 
\begin{equation}
\begin{split}
    \alpha_{\rm DM} &< 1.3 \times 10^{-16} \left(\frac{m_\chi}{10 \text{ eV}}\right)\left( \frac{1 \text{ cm}}{\Delta x}\right)^{\tfrac{1}{2}} \left(\frac{87}{A}\right)\\
    & \hspace{-0.075\linewidth}  \times   \left(\frac{0.1 \text{ s}}{\mathcal{T}_{1/2}}\right)^{\tfrac{1}{2}} \left(\frac{10}{\sf{L}}\right)^{\tfrac{1}{2}} \left(\frac{10^6}{N}\right)^{\tfrac{3}{4}} \left(\frac{1}{\aleph}\right)^{\tfrac{1}{4}} \left(\frac{1}{f_\chi}\right)^{1/2}\!\!\!\!,    \end{split}
\end{equation}
for $m_\phi r_c \ll 1$, where ${\sf L} = \log([m_\phi r_c]^{-2})$. For $m_\phi r_c \gg 1$, we get instead 
\begin{equation}
    \begin{split}
     \alpha_{\rm DM} &< 2.3 \times 10^{-14} \left(\frac{m_\chi}{10 \text{ eV}}\right) \left(\frac{1 \text{ cm}}{\Delta x}\right)^{\tfrac{1}{2}} \left(\frac{87}{A}\right)  \\ 
     & \hspace{-0.075\linewidth} \times    \left( \frac{0.1 \text{ s}}{\mathcal{T}_{1/2}}\right)^{\tfrac{1}{2}} \left(\frac{m_\phi r_c}{10 }\right)^2 \left(\frac{10^6}{N}\right)^{\tfrac{3}{4}} \left(\frac{1}{\aleph}\right)^{\tfrac{1}{4}}\left(\frac{1}{f_\chi}\right)^{1/2}\!\!\!\!.
    \end{split}
\end{equation}
In both cases we have assumed that the interferometer is aligned with the dark matter wind. 
We note that, for the benchmarks considered, \cref{eq:argsin} is satisfied and the analytic approximations are reliable as can be seen explicitly by comparing to numerical integration in  \cref{fig:plot}.

These results should be compared to complimentary sources of constraints, namely those from fifth forces searches (for the force induced between nucleons by the scalar $\phi$), self-interacting dark matter constraints from the bullet cluster (from the force between dark matter particles induced by $\phi$), and stellar cooling constraints (from the emission of $\phi$ quanta in stars). Fifth-force searches and stellar cooling constrain the combination $\alpha_n=g_n^2/(4\pi)$, whereas the bullet cluster constraints $\alpha_\chi= g_\chi^2/(4\pi)$. The latter demand that $\sigma_T/m_\chi \lesssim 1~{\rm cm}^2/g$ with $\sigma_T$ the momentum-transfer cross section. The bullet cluster constraint does not apply for sub-populations of dark matter (roughly below fractions of $10^{-2}$).

Our results are summarized in Fig.~\ref{fig:plot}, where one can see the competition between fifth-force constraints, stellar cooling constraints, and the atom interferometer observables shown here. We fix the fraction of dark matter $f_\chi=0.05$, for which bullet cluster constraints do not apply, fix the dark matter coupling to the mediator to be $g_\chi=\sqrt{4\pi}$, and benchmark two light,e particle-like dark matter masses of $m_\chi=10~{\rm eV}$ and $m_\chi = 1~{\rm keV}$. We find that current atom interferometers can exceed fifth force constraints in a narrow range of $m_\phi$, but that they are not competitive with stellar cooling bounds. More optimistically, we find that for the large $g_\chi$ considered, future proposals such as the AEDGE interferometer will be able to probe regions of parameter space at smaller value of $m_\phi$ which are currently unconstrained by fifth force constraints if the experiment is shot-noise limited.

\newpage

\section{Discussion and conclusions}
When a background gas interacts with an atom interferometer, the soft scatterings between the gas and heavy atoms lead to phase shifts and contrast loss. When a large number of atoms are used in a single run of the experiment, these observables can receive enhancements due to coherent interference of the relevant scattering amplitudes. We have extended the existing formalism for point-like clouds \cite{Badurina:2024nge} to include realistic features such as finite temperature effects, cloud spreading, and a finite cloud form factor. The impact of these realistic features on phenomenology has also been investigated. 

\bigskip 

All finite temperature and structure dependent effects arise from a partial trace over the coordinate degrees of freedom of the cloud at the time of measurement. When cloud spreading is negligible, \cref{eq:expl-soln-semiclassical} provides a closed-form solution for the density matrix at the time of read-out. When cloud spreading is important, but the interactions between the atoms and the gas can be treated perturbatively, \cref{eq:perturbative-decoherence-result} gives the leading-order result including time-dependent corrections arising from the spreading of the atomic cloud. These two equations are our major results. While \cref{eq:expl-soln-semiclassical} generalizes previous results in the literature (e.g. point-like limit in~\cite{Badurina:2024nge}, phase-shift effects in~\cite{Du:2022ceh,Du:2023eae}, decoherence in matter interferometers in~\cite{Riedel:2012ur,Riedel:2016acj}), \cref{eq:perturbative-decoherence-result} incorporates, for the first time, the influence of cloud spreading into the evolution formalism of an atom interferometer in a dilute, homogeneous, Markovian background gas. As we show explicitly for a thermal cloud, to leading order in perturbation theory, these effects are properly captured by a time-dependent cloud radius in agreement with basic physical intuition.

We have applied this formalism to an atom interferometer whose initial state is prepared at temperature $T$ in a harmonic trap. Potential experimental architectures that could test the results of this paper have been proposed (including infrared photon- and $1~{\rm eV}$ electron-scattering), that we believe to be realistic for current and near term experiments. Explicit expressions for the coherent phase shift have been derived (with all integrals computed analytically) as well as the incoherent phase shift in the case of electron scattering. We have sketched how the static approximation of \cref{eq:expl-soln-semiclassical} may be immediately applied to matter interferometers.  A discussion of dark matter phenomenology, including new closed-form solutions for the phase shift from a Yukawa potential, has been provided.

Our work opens the door to a reliable inclusion of coherent enhancements in searches for dark matter and other beyond the Standard Model physics with atom interferometers. This includes both the enhancement of the signal rate discussed above, but also allows one to quantify the effects from Standard Model backgrounds as first discussed in~\cite{Du:2023eae}. A detailed understanding of both signal and background is crucial to any future search with atom interferometers including: measurements of gravitational waves, precision measurements of fundamental constants, and searches for physics beyond the Standard Model. 

The enhancements to a particle-like dark matter signal scale with the number of atoms in the cloud and can easily exceed six orders of magnitude with present architecture, and even exceed ten orders of magnitude for futuristic proposals; they must be properly understood to interpret future experimental data. In the phenomenological applications presented above we have only studied the coherent and incoherent contributions to the phase shift. Other observables, including those related to contrast loss and statistical fluctuations, will be studied in future work.

\subsection*{Acknowledgments}
We thank Leonardo Badurina for a careful reading of this manuscript and for his collaboration during previous work~\cite{Badurina:2024nge}. We thank the  Institut de F\'isica d'Altes Energies and the CERN Theory group for their hospitality during portions of this work. 
RP was supported during large portions of this project by the U.S. Department of Energy, Office of Science, Office of High Energy Physics under Award Number DE-SC0011632, by the Neutrino Theory Network under Award Number DEAC02-07CH11359, and by the Walter Burke Institute for Theoretical Physics. 

\vfill

\appendix
\section{Second order exponentiation
\label{app:2order} }
The leading correction for the exponentiation of the form factors in \cref{eq:exponentiation} up to ${\cal O}(g^4 \mathcal{T}^2)$ is given by
\begin{equation}
\begin{split}
    \delta^{(2)}& = - \sum_{i=1}^6  \int_{q,q'} \int_{t,t'} \bigg( \omega_D(\vb{q}) \omega_D(\vb{q}') \xi^{DD}_i \\
    &  \ \ + 2 \omega_D(\vb{q}) \omega_U(\vb{q}')\xi_i^{DU} + \omega_U(\vb{q}) \omega_U(\vb{q}')\xi_i^{UU}\bigg)~.
\end{split}
\end{equation}
The terms $\xi^{DD}_i$, $\xi^{DU}_i$ and $\xi^{UU}_i$ spoil the exponentiation of $\langle \lambda \rangle$. These arise from degeneracies among the particle indices\footnote{Notice that $\langle \lambda \lambda \rangle = \langle \lambda \rangle \langle \lambda \rangle$ only when $i \neq j \neq k \neq \ell$ and the initial state is uncorrelated.} and are defined as 
\begin{align}
\xi^{X X'}_1 &= \sum_{i\neq j \neq k} \Y_{ij}^{X} (\vb{q}) \Y_{jk}^{X'} (\vb{q'}) \ \mathcal{G}_1(\vb{q},\vb{q'})~,\\
\xi^{X X'}_2 &= \sum_{i \neq j \neq k} \Y_{ij}^{X} (\vb{q}) \Y_{kj}^{X'} (\vb{q'}) \ \mathcal{G}_2(\vb{q},\vb{q'})~,\\
\xi^{X X'}_3 &= \sum_{i \neq j \neq k} \Y_{ij}^{X} (\vb{q}) \Y_{ki}^{X'} (\vb{q'}) \ \mathcal{G}_3(\vb{q},\vb{q'})~,\\
\xi^{X X'}_4 &= \sum_{i \neq j \neq k} \Y_{ij}^{X} (\vb{q}) \Y_{ik}^{X'} (\vb{q'}) \ \mathcal{G}_4(\vb{q},\vb{q'})~,\\
\xi^{X X'}_5 &= \sum_{i \neq j} \Y_{ij}^{X} (\vb{q}) \Y_{ij}^{X'} (\vb{q}') \ \mathcal{G}_5(\vb{q},\vb{q'})~,\\
\xi^{X X'}_6 &= \sum_{i \neq j} \Y_{ij}^{X} (\vb{q}) \Y_{ji}^{X'} (\vb{q}') \ \mathcal{G}_6(\vb{q},\vb{q'})~,
\end{align}
where 
\begin{align}
\mathcal{G}_1 &\equiv \!\! \big[ F(\vb{q}) F(\vb{q}'-\vb{q}) F(-\vb{q}') - G(\vb{q}) G(\vb{q}') \big]~, \\
\mathcal{G}_2 &\equiv \!\! \big[ F(\vb{q}) F(-\vb{q}-\vb{q}')F(\vb{q}') - G(\vb{q}) G(\vb{q}') \big] ~, \\
\mathcal{G}_3 &\equiv \!\! \big[ F(-\vb{q}) F(\vb{q}') F(\vb{q}-\vb{q}') - G(\vb{q}) G(\vb{q}')\big]~,\\
\mathcal{G}_4 &\equiv \!\! \big[ F(-\vb{q}) F(-\vb{q'}) F(\vb{q}+\vb{q}') - G(\vb{q}) G(\vb{q'}) \big]~, \\
\mathcal{G}_5 &\equiv \!\! \big[ G(\vb{q}+\vb{q}') - G(\vb{q}) G(\vb{q}')\big]~,\\
\mathcal{G}_6 &\equiv \!\! \big[ G(\vb{q}-\vb{q}') - G(\vb{q}) G(\vb{q}')\big]~.
\end{align} 
Here, $F(\vb{q}) = \int_y \varrho_y  \, \e^{\iu \vb{q} \cdot \vb{y}_i}$ and $G(\vb{q})$ is defined in \cref{eq:G}.
In particular, the first four terms arise from the degeneracy of a single pair of indices, while the last two terms come from multiple degeneracies. 

For a generic matrix element of the density matrix, $\langle N_L' | \rho_r | N_L \rangle$, with $n = N_L-N_L'$ and a total of $N$ atoms, we can split each sum of the particle indices involved in three sets. For $n>0$, $i \subset [1,N_L -n]\equiv \cone$, $i \subset [N_L-n+1,N_L] \equiv \ctwo$, and $i \subset [N_L+1,N] \equiv \cthree$. The cardinality of each set is:  $|\cone | = N_L-n$, $|\ctwo| = n$, and  $|\cthree| = N-N_L$. For $n<0$, we have instead $\cone = N_L$, $\ctwo =|n|$ and $\cthree = N - N_L - |n|$. 
Such a splitting is convenient as indices belonging to the same block lead to a degenerate result of $\Y^{X}_{ij}$, what allows us to write $\Y^{X}_{ij}$ in a block-matrix form: 
\begin{align}
\Y^D_{[i][j]} \!&\equiv \!\! \begin{pmatrix} \Y^D_{\cone \cone} & \Y^D_{\cone \ctwo} & \Y^D_{\cone \cthree} \\
\Y^D_{\ctwo \cone} & \Y^D_{\ctwo \ctwo} & \Y^D_{\ctwo \cthree} \\
\Y^D_{\cthree \cone} & \Y^D_{\cthree \ctwo} & \Y^D_{\cthree \cthree} \end{pmatrix} =  \begin{pmatrix} 0 & \triangle & 0 \\ -\triangle^* & \square & \triangle \\ 0 & -\triangle^* & 0 \end{pmatrix}, \qquad \label{eq:LambdaD}\\
\Y^U_{[i][j]} \! &\equiv \!\! \begin{pmatrix} \Y^U_{\cone \cone} & \Y^U_{\cone \ctwo} & \Y^U_{\cone \cthree} \\ \Y^U_{\ctwo \cone} & \Y^U_{\ctwo \ctwo} & \Y^U_{\ctwo \cthree} \\ \Y^U_{\cthree \cone} & \Y^U_{\cthree \ctwo} & \Y^U_{\cthree \cthree} \end{pmatrix} =  \begin{pmatrix} 0 & -\triangle & 0 \\
-\triangle^* & 0 & \triangle\\ 0 & \triangle^* & 0 \end{pmatrix},  \qquad \label{eq:LambdaU}
\end{align}
with $\triangle (\vb{q})= \tfrac{1}{2}(1-\e^{\iu \vb{q}\cdot \vb{\Delta x}})$ for $n>0$, and $\triangle (\vb{q}) = -\tfrac{1}{2}(1-\e^{\iu\vb{q}\cdot \vb{\Delta x}})$ for $n<0$. By inspection, for different sets $[i]\neq [j]$, 
\begin{align}
\Y_{[i][j]}^D (\vb{q}) &= - \Y_{[j][i]}^D(\vb{q})^* = -\Y_{[j][i]}^D(-\vb{q})~, \label{eq:antihermitian}\\ 
\Y_{[i][j]}^U(\vb{q}) &= \Y_{[j][i]}^U (\vb{q})^*~. \label{eq:hermitian}
\end{align}
Let us start with $\xi^{DD}$.
Applying the change of variable $\vb{q}' \to -\vb{q}'$, 
\begin{equation}
    \begin{split}
        \mathcal{G}_2 &\to \mathcal{G}_1~, \
        \mathcal{G}_4 \to \mathcal{G}_3~, \ \text{and}\
        \mathcal{G}_6 \to \mathcal{G}_5~.
    \end{split}
\end{equation}
Using Eq.~\eqref{eq:antihermitian}, and substituting the zeroes in $\Y_{[i][j]}^D$ we find that 
\begin{align*}
&\xi_1^{DD} + \xi_2^{DD}(\vb{q}' \to - \vb{q'}) = 2 \text{Re}[\Y_{\ctwo \ctwo}^D (\vb{q}')] |\ctwo|(|\ctwo|-1) \\
& \qquad \times \bigg[ |\cone| \Y^D_{\cone \ctwo}(\vb{q})  + |\cthree| \Y_{\cthree \ctwo}^D(\vb{q}) + (|\ctwo|-2) \Y^D_{\ctwo \ctwo}(\vb{q}) \bigg] \mathcal{G}_1~, \quad \nonumber\\
&\xi_3^{DD} + \xi_4^{DD}(\vb{q}' \to - \vb{q'}) = 2 \text{Re}[\Y_{\ctwo \ctwo}^D(\vb{q}')]  |\ctwo|(|\ctwo|-1) \\
& \qquad \times \bigg[|\cone| \Y^D_{\ctwo \cone}(\vb{q})  +|\cthree| \Y^D_{\ctwo \cthree} (\vb{q}) + (|\ctwo|-2) \Y^D_{\ctwo \ctwo}(\vb{q}) \bigg]\mathcal{G}_1^*~,\quad \nonumber\\
&\xi_5^{DD} + \xi_6^{DD}(\vb{q}' \to - \vb{q}') =\\
&\qquad \qquad  \qquad \ \, 2 \text{Re}[\Y_{\ctwo \ctwo}^D(\vb{q}')] |\ctwo|(|\ctwo|-1) \Y^D_{\ctwo \ctwo}(\vb{q}) \ \mathcal{G}_5~,\nonumber
\end{align*}
where we have used that $\mathcal{G}_3 = \mathcal{G}_1^*$.
By adding them under the change of variables $\xi_1^{DD} + \xi_2^{DD} + \big(\xi_3^{DD} +\xi_4^{DD})(\vb{q} \to - \vb{q} , \vb{q}' \to - \vb{q'}) \big)  + \tfrac{1}{2}\big(\xi_5^{DD} + \xi_6^{DD} \big) + \tfrac{1}{2}\big(\xi_5^{DD} + \xi_6^{DD})(\vb{q}\to -\vb{q}, \vb{q}'\to -\vb{q}'\big)$, we get that 
\begin{equation}
\begin{split}
\sum_{i=1}^6 \xi_i^{DD}&=  \frac{4|\ctwo| !}{  (|\ctwo|-2)!} \bigg(  ( |\ctwo|-2) \text{Re}[\mathcal{G}_1 ]+ \mathcal{G}_5 \bigg)\\
&\qquad  \times \text{Re}[\Y^D_{\ctwo \ctwo}(\vb{q})]\text{Re}[\Y^D_{\ctwo \ctwo}(\vb{q'})]~.
\end{split}
\end{equation}

For the interference term, $\xi^{DU} = \xi^{UD}$, performing the change of variable $\vb{q}\to -\vb{q}$, 
\begin{align}
&&\xi_1^{DU} + \xi_4^{DU}(\vb{q}\to - \vb{q}) = 2 \text{Re}[\Y_{\ctwo \ctwo}^D(\vb{q})] |\ctwo|(|\ctwo|-1) \nonumber \quad \\
&& \hspace{2cm} \times \bigg[ |\cone| \Y_{\ctwo \cone}^U(\vb{q}')+|\cthree| \Y_{\ctwo \cthree}^U(\vb{q}') \bigg]\mathcal{G}_1~, \quad\\
&&\xi_2^{DU} + \xi_3^{DU}(\vb{q} \to - \vb{q})= 2 \text{Re}[\Y_{\ctwo  \ctwo}^D(\vb{q})] |\ctwo|(|\ctwo |-1) \nonumber \quad \\
&& \hspace{2cm} \times \bigg[ |\cone| \Y_{\cone \ctwo}^U(\vb{q}') + |\cthree| \Y_{\cthree \ctwo}^U (\vb{q}') \bigg]\mathcal{G}_2~,\quad\\
&& \xi_5^{DU} + \xi_6^{DU}(\vb{q} \to -\vb{q})   = 0~.
\end{align}

Using Eq.~\eqref{eq:hermitian} we get 
\begin{equation}\label{eq:xiDU}
\begin{split}
\sum_{i=1}^6 \xi^{DU}_i& = \frac{4  |\ctwo|!}{ (|\ctwo|-2)!} \text{Re}[\Y_{\ctwo \ctwo}^D  (\vb{q}) ]  \\
&\ \times \text{Re} \bigg[ \bigg( |\cone| \Y_{\ctwo \cone}^U (\vb{q}') + |\cthree|  \Y_{\ctwo \cthree}^U(\vb{q})  \bigg) \mathcal{G}_1 \bigg]~.
\end{split}
\end{equation}
Finally, for $\xi^{UU}$, we have
\begin{widetext}
\begin{align}
 &\sum_{i=1}^6 \xi_i^{UU} = 2|\ctwo| \text{Re} \bigg[ |\cone| \Y^U_{\cone \ctwo}(\vb{q}) \bigg( \Y^U_{\cone \ctwo }(\vb{q}') \mathcal{G}_5 + \Y^U_{\ctwo \cone}(\vb{q}') \mathcal{G}_6 \bigg)  + |\cthree| \Y_{\ctwo \cthree}^U(\vb{q}) \bigg( \Y^U_{\ctwo \cthree}(\vb{q}') \mathcal{G}_5 + \Y^U_{\cthree \ctwo}(\vb{q}') \mathcal{G}_6 \bigg)  \bigg]\\
& + 2|\ctwo|\text{Re}  \bigg[  |\cone| \bigg((|\cone|-1)  \Y^U_{\cone \ctwo}(\vb{q}) +|\cthree|  \Y_{\cthree \ctwo}^U(\vb{q}) \bigg)\bigg( \Y^U_{\ctwo \cone}(\vb{q}') \mathcal{G}_1 + \Y^U_{\cone \ctwo}(\vb{q}') \mathcal{G}_2 \bigg)+ (|\ctwo|-1) |\cone| \Y^U_{\ctwo \cone}(\vb{q}) \big( \Y_{\cone \ctwo}^U(\vb{q}') \mathcal{G}_1 + \Y_{\ctwo \cone}^U(\vb{q}') \mathcal{G}_2 \big)\nonumber \\
&~~  + (|\ctwo|-1) |\cthree| \Y_{\ctwo \cthree}^U(\vb{q}) \big(\Y_{\cthree \ctwo}^U(\vb{q}') \mathcal{G}_1 + \Y_{\ctwo \cthree}^U(\vb{q}') \mathcal{G}_2 \big) + |\cthree| \bigg( (|\cthree|-1) \Y_{\cthree \ctwo}^U(\vb{q}) + |\cone| \Y_{\cone  \ctwo}^U(\vb{q}) \bigg)  \bigg( \Y^U_{\ctwo \cthree}(\vb{q}') \mathcal{G}_1 + \Y^U_{\cthree \ctwo}(\vb{q}') \mathcal{G}_2 \bigg) \bigg]~.\nonumber
\end{align}
\end{widetext}
This analysis explicitly demonstrates that, even for uncorrelated initial states, terms with coincident indices spoil the exponentiation of the first order terms. When considering an atom interferometer, for which correlations can be neglected, these violations of exponentiation are down by $1/N$ or more relative to the leading-order effects (the series in $g^2 N$). Therefore, up to corrections of $1/N$ these higher order terms may be neglected when convenient for $N\gg 1$.

\section{Analysis of integrals}
In this appendix we derive some of the analytic approximations to the integrals used above. 
\subsection{Asymptotic analysis for \cref{tricky-integral} \label{app:tricky-integral}}
We have used an asymptotic expansion of the following integral with $a$, $b$, and $\TT$ assumed real,
\begin{equation}\label{tricky-integral-def}
    I(a,b,\TT)=\int_{-\pi}^{\pi} \frac{\dd \phi}{2\pi} \int_0^{\TT}
    \!\!\dd t \sin(a t + b t \cos\phi )~.
\end{equation}
This formula is a little subtle and so we give details here for completeness. First we do the integral over time, 
\begin{equation}
    I(a,b,\TT)=\int_{-\pi}^{\pi} \frac{\dd \phi}{2\pi}  \, 
    \frac{1-\cos\qty(~\TT f(\phi)~)}{f(\phi)}~,
\end{equation}
with $f(\phi) = a+b\cos\phi$. Let us first consider the case $|a|>|b|$, for which the large-$\TT$ analysis of this integral is straightforward: one applies the Riemann-Lebesgue lemma (the cosine term is $1/\TT$ suppressed), and the integral gives 
\begin{equation}\label{eq:abigT}
    I(a,b,\TT) = \frac{{\rm sgn}(a)}{\sqrt{a^2-b^2}}  + O(1/\TT)~~~\text{for}~~~|a|>|b|~, 
\end{equation}
with ${\rm sgn}(a)= a/|a|$. For completeness, we include here the solution of $I(a,b,\TT)$ valid for any $\TT$, 
\begin{equation}\label{eq:Icomplete}
\begin{split}
 I(a,b,\TT) &= \frac{\text{sgn}(a)}{\sqrt{a^2 -  b^2}} \bigg( 1 - J_0(b \TT) \cos(a \TT)  \\
 &- 2\cos (a \TT) \sum_{k=0}^\infty (-)^k J_{2k}(b\TT) \kappa^{2k} \\
 & - 2 \sin(a \TT) \sum_{k=1}^\infty(-)^k J_{2k+1}(b\TT) \kappa^{2k+1}\bigg)~,
 \end{split}
\end{equation}
with 
\begin{equation}
\kappa =\left(\frac{a-\sqrt{a^2-b^2}}{b}\right).
\end{equation}
\Cref{eq:Icomplete} maps to \cref{eq:abigT} in the limit $\TT \to \infty$.

Next consider $|a|<|b|$. Now the function $f(\phi)$ has roots at two values of $\phi$ in the integration domain. Since $\cos\phi$ is even in $\phi$, we may equivalently study 
\begin{equation}
     I(a,b,\TT) = \int_{0}^{\pi} \frac{\dd \phi}{\pi} \, 
    \frac{1-\cos\qty(~\TT f(\phi)~)}{f(\phi)}~.
\end{equation}
In this case, we cannot directly apply the Riemann-Lebesgue lemma, since $1/f(\phi)$ becomes very large near the root of $f(\phi)$ at $\phi=\arccos(-a/b)$. 

Let us proceed by changing variables to $u= \TT f(\phi)$, in terms of which we have 
\begin{equation}
    I(a,b,\TT) = \!\!\int_{(a-b) \TT}^{(a+b)\TT} \!\! \dd u \frac{1}{\sqrt{b^2 -(a-\tfrac{u}{\TT})^2}} \frac{1-\cos u}{u} ~, 
\end{equation}
with the root of $f$ having been mapped to $u=0$. If we now break up the integration according to 
\begin{equation}
    \int_{(a-b) \TT}^{(b+a)\TT}\!\! \dd u = \int_{(a-b) \TT}^{-\Delta}\!\! \dd u ~+~\int_{-\Delta}^{\Delta}\!\! \dd u ~+~\int_{\Delta}^{(b+a)\TT}\!\! \dd u~,
\end{equation}
with $\Delta\sim O(1)$, 
then one can easily show that the middle integral is $O(\Delta/\TT)$ and can be dropped. By changing variables, we then have
\begin{equation}
    \begin{split}
    I(a,b,\TT) &= \!\!\int_{\Delta}^{(b+a)\TT}\!\! \dd u \frac{1}{\sqrt{b^2 -(a-\tfrac{u}{\TT})^2}} \frac{1-\cos u}{u} ~\\
    &- \!\!\int_{\Delta}^{(b-a)\TT}\!\! \dd u \frac{1}{\sqrt{b^2 -(a+\tfrac{u}{\TT})^2}} \frac{1-\cos u}{u}~ . 
    \end{split}
\end{equation}
Now the difference between the two integrands near $u\sim \Delta$ results in a contribution that is $O(1/\TT)$ suppressed. This allows us to drop the $\cos u$ term by the Riemann-Lebesgue lemma (notice that this is $\cos u= \cos \TT f$ with $f\sim O(1)$ and therefore rapidly oscillating), and we find, up to $1/\TT$ suppressed terms, 
\begin{equation}
    \begin{split}
    I(a,b,\TT) &= \!\!\int_{\Delta}^{(b+a)\TT}\!\! \dd u \frac{1}{u\sqrt{b^2 -(a-\tfrac{u}{\TT})^2}}  ~\\
    &- \!\!\int_{\Delta}^{(b-a)\TT}\!\! \dd u \frac{1}{u\sqrt{b^2 -(a+\tfrac{u}{\TT})^2}} ~ . 
    \end{split}
\end{equation}
A change of variables $b\rightarrow -b$ and $\TT \rightarrow -\TT$ on the second integral then immediately yields 
\begin{equation}
      I(a,b,\TT) = 0 + O\qty(\frac{1}{\TT})~~~\text{for}~~ a<b~. 
\end{equation}

\subsection{Asymptotic analysis for \cref{phase-shift-LRDM} \label{app:LRDM-integral}}
To compute \cref{phase-shift-LRDM} we begin from the following integral,
\begin{equation}
    I_q = \int \dd q \, \frac{q^2}{(q^2+ m_\phi^2)^2}J_1(q \Delta x_\perp) \e^{-\tfrac{1}{2}r_c^2 q^2} ~,
\end{equation}
where we have performed the time integral in \cref{eq:gammachi} by applying \cref{eq:timeintegral}.
We introduce an auxiliary scale $\mathcal{Q}$ satisfying the hierarchy $m_\phi \ll \mathcal{Q} \ll 1/\Delta x_\perp$, and split the above integral accordingly, i.e. $I_q = I_{\text{low}\, q} + I_{\text{high}\, q}$. In the region $0 \leq q \leq \mathcal{Q}$, 
\begin{equation}
\begin{split}
 I_{\text{low}\,q} &\simeq \frac{\Delta x_\perp}{2} \int_0^{\mathcal{Q}} \dd q \frac{q^3}{(q^2 + m_\phi^2)^2}\\
 & \simeq \frac{\Delta x_\perp}{4} (\log \mathcal{Q}^2 - \log m_\phi^2 - 1)~,
 \end{split}
\end{equation}
where we have expanded $J_1 (q \Delta x_\perp) \simeq q \Delta x_\perp /2$, and taken $G(\vb{q})\simeq 1$.
In the region $\mathcal{Q} \leq q \ll p_\chi$, we can Taylor-expand the propagator in $m_\phi^2/q^2$, obtaining 
\begin{align}
        I_{\text{high}\, q} &\simeq \int_{\mathcal{Q}}^{1/r_c} \frac{\dd q}{q^2} J_1(q \Delta x_\perp) \nonumber \\
        &\simeq - \frac{\Delta x_\perp}{4}  \log \qty(\mathcal{Q}^2 r_c^2) + \ldots  ~. \nonumber
    \end{align}
Above, we have changed the Gaussian cloud form factor into a step function with a maximum momentum transfer at $1/r_c$. This reproduces the logarithmic term reliably, but does not properly capture finite pieces for the integral (which we therefore write as ellipses). Adding both contributions and keeping only the logarithmic term we arrive at \cref{phase-shift-LRDM} with the dependence on $\mathcal{Q}$ canceling as expected.

\bibliography{biblio.bib}

\begin{thebibliography}{54}%
\makeatletter
\providecommand \@ifxundefined [1]{%
 \@ifx{#1\undefined}
}%
\providecommand \@ifnum [1]{%
 \ifnum #1\expandafter \@firstoftwo
 \else \expandafter \@secondoftwo
 \fi
}%
\providecommand \@ifx [1]{%
 \ifx #1\expandafter \@firstoftwo
 \else \expandafter \@secondoftwo
 \fi
}%
\providecommand \natexlab [1]{#1}%
\providecommand \enquote  [1]{``#1''}%
\providecommand \bibnamefont  [1]{#1}%
\providecommand \bibfnamefont [1]{#1}%
\providecommand \citenamefont [1]{#1}%
\providecommand \href@noop [0]{\@secondoftwo}%
\providecommand \href [0]{\begingroup \@sanitize@url \@href}%
\providecommand \@href[1]{\@@startlink{#1}\@@href}%
\providecommand \@@href[1]{\endgroup#1\@@endlink}%
\providecommand \@sanitize@url [0]{\catcode `\\12\catcode `\$12\catcode
  `\&12\catcode `\#12\catcode `\^12\catcode `\_12\catcode `\%12\relax}%
\providecommand \@@startlink[1]{}%
\providecommand \@@endlink[0]{}%
\providecommand \url  [0]{\begingroup\@sanitize@url \@url }%
\providecommand \@url [1]{\endgroup\@href {#1}{\urlprefix }}%
\providecommand \urlprefix  [0]{URL }%
\providecommand \Eprint [0]{\href }%
\providecommand \doibase [0]{http://dx.doi.org/}%
\providecommand \selectlanguage [0]{\@gobble}%
\providecommand \bibinfo  [0]{\@secondoftwo}%
\providecommand \bibfield  [0]{\@secondoftwo}%
\providecommand \translation [1]{[#1]}%
\providecommand \BibitemOpen [0]{}%
\providecommand \bibitemStop [0]{}%
\providecommand \bibitemNoStop [0]{.\EOS\space}%
\providecommand \EOS [0]{\spacefactor3000\relax}%
\providecommand \BibitemShut  [1]{\csname bibitem#1\endcsname}%
\let\auto@bib@innerbib\@empty
\bibitem [{\citenamefont {Kasevich}\ and\ \citenamefont
  {Chu}(1992)}]{Kasevich:1992yii}%
  \BibitemOpen
  \bibfield  {author} {\bibinfo {author} {\bibfnamefont {M.}~\bibnamefont
  {Kasevich}}\ and\ \bibinfo {author} {\bibfnamefont {S.}~\bibnamefont {Chu}},\
  }\bibfield  {title} {\enquote {\bibinfo {title} {{Measurement of the
  gravitational acceleration of an atom with a light-pulse atom
  interferometer}},}\ }\href {\doibase 10.1007/BF00325375} {\bibfield
  {journal} {\bibinfo  {journal} {Appl. Phys. B}\ }\textbf {\bibinfo {volume}
  {54}},\ \bibinfo {pages} {321--332} (\bibinfo {year} {1992})}\BibitemShut
  {NoStop}%
\bibitem [{\citenamefont {Zhou}\ \emph {et~al.}(2015)\citenamefont {Zhou} \emph
  {et~al.}}]{Zhou:2015pna}%
  \BibitemOpen
  \bibfield  {author} {\bibinfo {author} {\bibfnamefont {Lin}\ \bibnamefont
  {Zhou}} \emph {et~al.},\ }\bibfield  {title} {\enquote {\bibinfo {title}
  {{Test of Equivalence Principle at $10^{-8}$ Level by a Dual-species
  Double-diffraction Raman Atom Interferometer}},}\ }\href {\doibase
  10.1103/PhysRevLett.115.013004} {\bibfield  {journal} {\bibinfo  {journal}
  {Phys. Rev. Lett.}\ }\textbf {\bibinfo {volume} {115}},\ \bibinfo {pages}
  {013004} (\bibinfo {year} {2015})},\ \Eprint
  {http://arxiv.org/abs/1503.00401} {arXiv:1503.00401 [physics.atom-ph]}
  \BibitemShut {NoStop}%
\bibitem [{\citenamefont {Rosi}\ \emph {et~al.}(2017)\citenamefont {Rosi},
  \citenamefont {D'Amico}, \citenamefont {Cacciapuoti}, \citenamefont
  {Sorrentino}, \citenamefont {Prevedelli}, \citenamefont {Zych}, \citenamefont
  {Brukner},\ and\ \citenamefont {Tino}}]{Rosi:2017ieh}%
  \BibitemOpen
  \bibfield  {author} {\bibinfo {author} {\bibfnamefont {G.}~\bibnamefont
  {Rosi}}, \bibinfo {author} {\bibfnamefont {G.}~\bibnamefont {D'Amico}},
  \bibinfo {author} {\bibfnamefont {L.}~\bibnamefont {Cacciapuoti}}, \bibinfo
  {author} {\bibfnamefont {F.}~\bibnamefont {Sorrentino}}, \bibinfo {author}
  {\bibfnamefont {M.}~\bibnamefont {Prevedelli}}, \bibinfo {author}
  {\bibfnamefont {M.}~\bibnamefont {Zych}}, \bibinfo {author} {\bibfnamefont
  {C.}~\bibnamefont {Brukner}}, \ and\ \bibinfo {author} {\bibfnamefont
  {G.~M.}\ \bibnamefont {Tino}},\ }\bibfield  {title} {\enquote {\bibinfo
  {title} {{Quantum test of the equivalence principle for atoms in
  superpositions of internal energy eigenstates}},}\ }\href {\doibase
  10.1038/ncomms15529} {\bibfield  {journal} {\bibinfo  {journal} {Nature
  Commun.}\ }\textbf {\bibinfo {volume} {8}},\ \bibinfo {pages} {5529}
  (\bibinfo {year} {2017})},\ \Eprint {http://arxiv.org/abs/1704.02296}
  {arXiv:1704.02296 [physics.atom-ph]} \BibitemShut {NoStop}%
\bibitem [{\citenamefont {Yu}\ \emph {et~al.}(2019)\citenamefont {Yu},
  \citenamefont {Zhong}, \citenamefont {Estey}, \citenamefont {Kwan},
  \citenamefont {Parker},\ and\ \citenamefont {M\"uller}}]{Yu:2019gdh}%
  \BibitemOpen
  \bibfield  {author} {\bibinfo {author} {\bibfnamefont {Chenghui}\
  \bibnamefont {Yu}}, \bibinfo {author} {\bibfnamefont {Weicheng}\ \bibnamefont
  {Zhong}}, \bibinfo {author} {\bibfnamefont {Brian}\ \bibnamefont {Estey}},
  \bibinfo {author} {\bibfnamefont {Joyce}\ \bibnamefont {Kwan}}, \bibinfo
  {author} {\bibfnamefont {Richard~H.}\ \bibnamefont {Parker}}, \ and\ \bibinfo
  {author} {\bibfnamefont {Holger}\ \bibnamefont {M\"uller}},\ }\bibfield
  {title} {\enquote {\bibinfo {title} {{Atom-Interferometry Measurement of the
  Fine Structure Constant}},}\ }\href {\doibase 10.1002/andp.201800346}
  {\bibfield  {journal} {\bibinfo  {journal} {Annalen Phys.}\ }\textbf
  {\bibinfo {volume} {531}},\ \bibinfo {pages} {1800346} (\bibinfo {year}
  {2019})}\BibitemShut {NoStop}%
\bibitem [{\citenamefont {Morel}\ \emph {et~al.}(2020)\citenamefont {Morel},
  \citenamefont {Yao}, \citenamefont {Clad\'e},\ and\ \citenamefont
  {Guellati-Kh\'elifa}}]{Morel:2020dww}%
  \BibitemOpen
  \bibfield  {author} {\bibinfo {author} {\bibfnamefont {L\'eo}\ \bibnamefont
  {Morel}}, \bibinfo {author} {\bibfnamefont {Zhibin}\ \bibnamefont {Yao}},
  \bibinfo {author} {\bibfnamefont {Pierre}\ \bibnamefont {Clad\'e}}, \ and\
  \bibinfo {author} {\bibfnamefont {Sa\"\i{}da}\ \bibnamefont
  {Guellati-Kh\'elifa}},\ }\bibfield  {title} {\enquote {\bibinfo {title}
  {{Determination of the fine-structure constant with an accuracy of 81 parts
  per trillion}},}\ }\href {\doibase 10.1038/s41586-020-2964-7} {\bibfield
  {journal} {\bibinfo  {journal} {Nature}\ }\textbf {\bibinfo {volume} {588}},\
  \bibinfo {pages} {61--65} (\bibinfo {year} {2020})}\BibitemShut {NoStop}%
\bibitem [{\citenamefont {Arvanitaki}\ \emph {et~al.}(2018)\citenamefont
  {Arvanitaki}, \citenamefont {Graham}, \citenamefont {Hogan}, \citenamefont
  {Rajendran},\ and\ \citenamefont {Van~Tilburg}}]{Arvanitaki:2016fyj}%
  \BibitemOpen
  \bibfield  {author} {\bibinfo {author} {\bibfnamefont {Asimina}\ \bibnamefont
  {Arvanitaki}}, \bibinfo {author} {\bibfnamefont {Peter~W.}\ \bibnamefont
  {Graham}}, \bibinfo {author} {\bibfnamefont {Jason~M.}\ \bibnamefont
  {Hogan}}, \bibinfo {author} {\bibfnamefont {Surjeet}\ \bibnamefont
  {Rajendran}}, \ and\ \bibinfo {author} {\bibfnamefont {Ken}\ \bibnamefont
  {Van~Tilburg}},\ }\bibfield  {title} {\enquote {\bibinfo {title} {{Search for
  light scalar dark matter with atomic gravitational wave detectors}},}\ }\href
  {\doibase 10.1103/PhysRevD.97.075020} {\bibfield  {journal} {\bibinfo
  {journal} {Phys. Rev. D}\ }\textbf {\bibinfo {volume} {97}},\ \bibinfo
  {pages} {075020} (\bibinfo {year} {2018})},\ \Eprint
  {http://arxiv.org/abs/1606.04541} {arXiv:1606.04541 [hep-ph]} \BibitemShut
  {NoStop}%
\bibitem [{\citenamefont {Geraci}\ and\ \citenamefont
  {Derevianko}(2016)}]{Geraci:2016fva}%
  \BibitemOpen
  \bibfield  {author} {\bibinfo {author} {\bibfnamefont {Andrew~A.}\
  \bibnamefont {Geraci}}\ and\ \bibinfo {author} {\bibfnamefont {Andrei}\
  \bibnamefont {Derevianko}},\ }\bibfield  {title} {\enquote {\bibinfo {title}
  {{Sensitivity of atom interferometry to ultralight scalar field dark
  matter}},}\ }\href {\doibase 10.1103/PhysRevLett.117.261301} {\bibfield
  {journal} {\bibinfo  {journal} {Phys. Rev. Lett.}\ }\textbf {\bibinfo
  {volume} {117}},\ \bibinfo {pages} {261301} (\bibinfo {year} {2016})},\
  \Eprint {http://arxiv.org/abs/1605.04048} {arXiv:1605.04048
  [physics.atom-ph]} \BibitemShut {NoStop}%
\bibitem [{\citenamefont {Blas}\ \emph {et~al.}(2025)\citenamefont {Blas},
  \citenamefont {Carlton},\ and\ \citenamefont {McCabe}}]{Blas:2024kps}%
  \BibitemOpen
  \bibfield  {author} {\bibinfo {author} {\bibfnamefont {Diego}\ \bibnamefont
  {Blas}}, \bibinfo {author} {\bibfnamefont {John}\ \bibnamefont {Carlton}}, \
  and\ \bibinfo {author} {\bibfnamefont {Christopher}\ \bibnamefont {McCabe}},\
  }\bibfield  {title} {\enquote {\bibinfo {title} {{Massive graviton dark
  matter searches with long-baseline atom interferometers}},}\ }\href {\doibase
  10.1103/zxtk-bwnf} {\bibfield  {journal} {\bibinfo  {journal} {Phys. Rev. D}\
  }\textbf {\bibinfo {volume} {111}},\ \bibinfo {pages} {115020} (\bibinfo
  {year} {2025})},\ \Eprint {http://arxiv.org/abs/2412.14282} {arXiv:2412.14282
  [hep-ph]} \BibitemShut {NoStop}%
\bibitem [{\citenamefont {Badurina}\ \emph {et~al.}(2025)\citenamefont
  {Badurina}, \citenamefont {Du}, \citenamefont {Lee}, \citenamefont {Wang},\
  and\ \citenamefont {Zurek}}]{Badurina:2025xwl}%
  \BibitemOpen
  \bibfield  {author} {\bibinfo {author} {\bibfnamefont {Leonardo}\
  \bibnamefont {Badurina}}, \bibinfo {author} {\bibfnamefont {Yufeng}\
  \bibnamefont {Du}}, \bibinfo {author} {\bibfnamefont {Vincent S.~H.}\
  \bibnamefont {Lee}}, \bibinfo {author} {\bibfnamefont {Yikun}\ \bibnamefont
  {Wang}}, \ and\ \bibinfo {author} {\bibfnamefont {Kathryn~M.}\ \bibnamefont
  {Zurek}},\ }\bibfield  {title} {\enquote {\bibinfo {title} {{Detecting
  gravitational signatures of dark matter with atom gradiometers}},}\ }\href
  {\doibase 10.1103/xs7b-zgtj} {\bibfield  {journal} {\bibinfo  {journal}
  {Phys. Rev. D}\ }\textbf {\bibinfo {volume} {112}},\ \bibinfo {pages}
  {063014} (\bibinfo {year} {2025})},\ \Eprint
  {http://arxiv.org/abs/2505.00781} {arXiv:2505.00781 [hep-ph]} \BibitemShut
  {NoStop}%
\bibitem [{\citenamefont {Dimopoulos}\ \emph {et~al.}(2009)\citenamefont
  {Dimopoulos}, \citenamefont {Graham}, \citenamefont {Hogan}, \citenamefont
  {Kasevich},\ and\ \citenamefont {Rajendran}}]{Dimopoulos:2007cj}%
  \BibitemOpen
  \bibfield  {author} {\bibinfo {author} {\bibfnamefont {Savas}\ \bibnamefont
  {Dimopoulos}}, \bibinfo {author} {\bibfnamefont {Peter~W.}\ \bibnamefont
  {Graham}}, \bibinfo {author} {\bibfnamefont {Jason~M.}\ \bibnamefont
  {Hogan}}, \bibinfo {author} {\bibfnamefont {Mark~A.}\ \bibnamefont
  {Kasevich}}, \ and\ \bibinfo {author} {\bibfnamefont {Surjeet}\ \bibnamefont
  {Rajendran}},\ }\bibfield  {title} {\enquote {\bibinfo {title}
  {{Gravitational Wave Detection with Atom Interferometry}},}\ }\href {\doibase
  10.1016/j.physletb.2009.06.011} {\bibfield  {journal} {\bibinfo  {journal}
  {Phys. Lett. B}\ }\textbf {\bibinfo {volume} {678}},\ \bibinfo {pages}
  {37--40} (\bibinfo {year} {2009})},\ \Eprint {http://arxiv.org/abs/0712.1250}
  {arXiv:0712.1250 [gr-qc]} \BibitemShut {NoStop}%
\bibitem [{\citenamefont {Graham}\ \emph {et~al.}(2013)\citenamefont {Graham},
  \citenamefont {Hogan}, \citenamefont {Kasevich},\ and\ \citenamefont
  {Rajendran}}]{Graham:2012sy}%
  \BibitemOpen
  \bibfield  {author} {\bibinfo {author} {\bibfnamefont {Peter~W.}\
  \bibnamefont {Graham}}, \bibinfo {author} {\bibfnamefont {Jason~M.}\
  \bibnamefont {Hogan}}, \bibinfo {author} {\bibfnamefont {Mark~A.}\
  \bibnamefont {Kasevich}}, \ and\ \bibinfo {author} {\bibfnamefont {Surjeet}\
  \bibnamefont {Rajendran}},\ }\bibfield  {title} {\enquote {\bibinfo {title}
  {{A New Method for Gravitational Wave Detection with Atomic Sensors}},}\
  }\href {\doibase 10.1103/PhysRevLett.110.171102} {\bibfield  {journal}
  {\bibinfo  {journal} {Phys. Rev. Lett.}\ }\textbf {\bibinfo {volume} {110}},\
  \bibinfo {pages} {171102} (\bibinfo {year} {2013})},\ \Eprint
  {http://arxiv.org/abs/1206.0818} {arXiv:1206.0818 [quant-ph]} \BibitemShut
  {NoStop}%
\bibitem [{\citenamefont {Graham}\ \emph
  {et~al.}(2016{\natexlab{a}})\citenamefont {Graham}, \citenamefont {Hogan},
  \citenamefont {Kasevich},\ and\ \citenamefont {Rajendran}}]{Graham:2016plp}%
  \BibitemOpen
  \bibfield  {author} {\bibinfo {author} {\bibfnamefont {Peter~W.}\
  \bibnamefont {Graham}}, \bibinfo {author} {\bibfnamefont {Jason~M.}\
  \bibnamefont {Hogan}}, \bibinfo {author} {\bibfnamefont {Mark~A.}\
  \bibnamefont {Kasevich}}, \ and\ \bibinfo {author} {\bibfnamefont {Surjeet}\
  \bibnamefont {Rajendran}},\ }\bibfield  {title} {\enquote {\bibinfo {title}
  {{Resonant mode for gravitational wave detectors based on atom
  interferometry}},}\ }\href {\doibase 10.1103/PhysRevD.94.104022} {\bibfield
  {journal} {\bibinfo  {journal} {Phys. Rev. D}\ }\textbf {\bibinfo {volume}
  {94}},\ \bibinfo {pages} {104022} (\bibinfo {year} {2016}{\natexlab{a}})},\
  \Eprint {http://arxiv.org/abs/1606.01860} {arXiv:1606.01860
  [physics.atom-ph]} \BibitemShut {NoStop}%
\bibitem [{\citenamefont {Wacker}(2010)}]{Wacker:2009ag}%
  \BibitemOpen
  \bibfield  {author} {\bibinfo {author} {\bibfnamefont {Jay~G.}\ \bibnamefont
  {Wacker}},\ }\bibfield  {title} {\enquote {\bibinfo {title} {{Using Atom
  Interferometry to Search for New Forces}},}\ }\href {\doibase
  10.1016/j.physletb.2010.04.072} {\bibfield  {journal} {\bibinfo  {journal}
  {Phys. Lett. B}\ }\textbf {\bibinfo {volume} {690}},\ \bibinfo {pages}
  {38--41} (\bibinfo {year} {2010})},\ \Eprint {http://arxiv.org/abs/0908.2447}
  {arXiv:0908.2447 [hep-ph]} \BibitemShut {NoStop}%
\bibitem [{\citenamefont {Graham}\ \emph
  {et~al.}(2016{\natexlab{b}})\citenamefont {Graham}, \citenamefont {Kaplan},
  \citenamefont {Mardon}, \citenamefont {Rajendran},\ and\ \citenamefont
  {Terrano}}]{Graham:2015ifn}%
  \BibitemOpen
  \bibfield  {author} {\bibinfo {author} {\bibfnamefont {Peter~W.}\
  \bibnamefont {Graham}}, \bibinfo {author} {\bibfnamefont {David~E.}\
  \bibnamefont {Kaplan}}, \bibinfo {author} {\bibfnamefont {Jeremy}\
  \bibnamefont {Mardon}}, \bibinfo {author} {\bibfnamefont {Surjeet}\
  \bibnamefont {Rajendran}}, \ and\ \bibinfo {author} {\bibfnamefont
  {William~A.}\ \bibnamefont {Terrano}},\ }\bibfield  {title} {\enquote
  {\bibinfo {title} {{Dark Matter Direct Detection with Accelerometers}},}\
  }\href {\doibase 10.1103/PhysRevD.93.075029} {\bibfield  {journal} {\bibinfo
  {journal} {Phys. Rev. D}\ }\textbf {\bibinfo {volume} {93}},\ \bibinfo
  {pages} {075029} (\bibinfo {year} {2016}{\natexlab{b}})},\ \Eprint
  {http://arxiv.org/abs/1512.06165} {arXiv:1512.06165 [hep-ph]} \BibitemShut
  {NoStop}%
\bibitem [{\citenamefont {Abe}\ \emph {et~al.}(2025)\citenamefont {Abe},
  \citenamefont {Hogan}, \citenamefont {Kaplan}, \citenamefont {Overstreet},\
  and\ \citenamefont {Rajendran}}]{Abe:2024idx}%
  \BibitemOpen
  \bibfield  {author} {\bibinfo {author} {\bibfnamefont {Mahiro}\ \bibnamefont
  {Abe}}, \bibinfo {author} {\bibfnamefont {Jason~M.}\ \bibnamefont {Hogan}},
  \bibinfo {author} {\bibfnamefont {David~E.}\ \bibnamefont {Kaplan}}, \bibinfo
  {author} {\bibfnamefont {Chris}\ \bibnamefont {Overstreet}}, \ and\ \bibinfo
  {author} {\bibfnamefont {Surjeet}\ \bibnamefont {Rajendran}},\ }\bibfield
  {title} {\enquote {\bibinfo {title} {{Search for monopole-dipole interactions
  with atom interferometry}},}\ }\href {\doibase 10.1103/vmkt-97vl} {\bibfield
  {journal} {\bibinfo  {journal} {Phys. Rev. D}\ }\textbf {\bibinfo {volume}
  {111}},\ \bibinfo {pages} {115009} (\bibinfo {year} {2025})},\ \Eprint
  {http://arxiv.org/abs/2409.14793} {arXiv:2409.14793 [hep-ph]} \BibitemShut
  {NoStop}%
\bibitem [{\citenamefont {Riedel}(2013)}]{Riedel:2012ur}%
  \BibitemOpen
  \bibfield  {author} {\bibinfo {author} {\bibfnamefont {C.~Jess}\ \bibnamefont
  {Riedel}},\ }\bibfield  {title} {\enquote {\bibinfo {title} {{Direct
  detection of classically undetectable dark matter through quantum
  decoherence}},}\ }\href {\doibase 10.1103/PhysRevD.88.116005} {\bibfield
  {journal} {\bibinfo  {journal} {Phys. Rev. D}\ }\textbf {\bibinfo {volume}
  {88}},\ \bibinfo {pages} {116005} (\bibinfo {year} {2013})},\ \Eprint
  {http://arxiv.org/abs/1212.3061} {arXiv:1212.3061 [quant-ph]} \BibitemShut
  {NoStop}%
\bibitem [{\citenamefont {Riedel}\ and\ \citenamefont
  {Yavin}(2017)}]{Riedel:2016acj}%
  \BibitemOpen
  \bibfield  {author} {\bibinfo {author} {\bibfnamefont {C.~Jess}\ \bibnamefont
  {Riedel}}\ and\ \bibinfo {author} {\bibfnamefont {Itay}\ \bibnamefont
  {Yavin}},\ }\bibfield  {title} {\enquote {\bibinfo {title} {{Decoherence as a
  way to measure extremely soft collisions with dark matter}},}\ }\href
  {\doibase 10.1103/PhysRevD.96.023007} {\bibfield  {journal} {\bibinfo
  {journal} {Phys. Rev. D}\ }\textbf {\bibinfo {volume} {96}},\ \bibinfo
  {pages} {023007} (\bibinfo {year} {2017})},\ \Eprint
  {http://arxiv.org/abs/1609.04145} {arXiv:1609.04145 [quant-ph]} \BibitemShut
  {NoStop}%
\bibitem [{\citenamefont {Du}\ \emph {et~al.}(2022)\citenamefont {Du},
  \citenamefont {Murgui}, \citenamefont {Pardo}, \citenamefont {Wang},\ and\
  \citenamefont {Zurek}}]{Du:2022ceh}%
  \BibitemOpen
  \bibfield  {author} {\bibinfo {author} {\bibfnamefont {Yufeng}\ \bibnamefont
  {Du}}, \bibinfo {author} {\bibfnamefont {Clara}\ \bibnamefont {Murgui}},
  \bibinfo {author} {\bibfnamefont {Kris}\ \bibnamefont {Pardo}}, \bibinfo
  {author} {\bibfnamefont {Yikun}\ \bibnamefont {Wang}}, \ and\ \bibinfo
  {author} {\bibfnamefont {Kathryn~M.}\ \bibnamefont {Zurek}},\ }\bibfield
  {title} {\enquote {\bibinfo {title} {{Atom interferometer tests of dark
  matter}},}\ }\href {\doibase 10.1103/PhysRevD.106.095041} {\bibfield
  {journal} {\bibinfo  {journal} {Phys. Rev. D}\ }\textbf {\bibinfo {volume}
  {106}},\ \bibinfo {pages} {095041} (\bibinfo {year} {2022})},\ \Eprint
  {http://arxiv.org/abs/2205.13546} {arXiv:2205.13546 [hep-ph]} \BibitemShut
  {NoStop}%
\bibitem [{\citenamefont {Joos}\ and\ \citenamefont {Zeh}(1985)}]{Joos:1984uk}%
  \BibitemOpen
  \bibfield  {author} {\bibinfo {author} {\bibfnamefont {E.}~\bibnamefont
  {Joos}}\ and\ \bibinfo {author} {\bibfnamefont {H.~D.}\ \bibnamefont {Zeh}},\
  }\bibfield  {title} {\enquote {\bibinfo {title} {{The Emergence of classical
  properties through interaction with the environment}},}\ }\href {\doibase
  10.1007/BF01725541} {\bibfield  {journal} {\bibinfo  {journal} {Z. Phys. B}\
  }\textbf {\bibinfo {volume} {59}},\ \bibinfo {pages} {223--243} (\bibinfo
  {year} {1985})}\BibitemShut {NoStop}%
\bibitem [{\citenamefont {Gallis}\ and\ \citenamefont
  {Fleming}(1990{\natexlab{a}})}]{PhysRevA.42.38}%
  \BibitemOpen
  \bibfield  {author} {\bibinfo {author} {\bibfnamefont {Michael~R.}\
  \bibnamefont {Gallis}}\ and\ \bibinfo {author} {\bibfnamefont {Gordon~N.}\
  \bibnamefont {Fleming}},\ }\bibfield  {title} {\enquote {\bibinfo {title}
  {Environmental and spontaneous localization},}\ }\href {\doibase
  10.1103/PhysRevA.42.38} {\bibfield  {journal} {\bibinfo  {journal} {Phys.
  Rev. A}\ }\textbf {\bibinfo {volume} {42}},\ \bibinfo {pages} {38--48}
  (\bibinfo {year} {1990}{\natexlab{a}})}\BibitemShut {NoStop}%
\bibitem [{\citenamefont {Hornberger}\ and\ \citenamefont
  {Sipe}(2003{\natexlab{a}})}]{Hornberger_2003}%
  \BibitemOpen
  \bibfield  {author} {\bibinfo {author} {\bibfnamefont {Klaus}\ \bibnamefont
  {Hornberger}}\ and\ \bibinfo {author} {\bibfnamefont {John~E.}\ \bibnamefont
  {Sipe}},\ }\bibfield  {title} {\enquote {\bibinfo {title} {Collisional
  decoherence reexamined},}\ }\href {\doibase 10.1103/physreva.68.012105}
  {\bibfield  {journal} {\bibinfo  {journal} {Physical Review A}\ }\textbf
  {\bibinfo {volume} {68}} (\bibinfo {year} {2003}{\natexlab{a}}),\
  10.1103/physreva.68.012105}\BibitemShut {NoStop}%
\bibitem [{\citenamefont {Tegmark}(1993)}]{Tegmark:1993yn}%
  \BibitemOpen
  \bibfield  {author} {\bibinfo {author} {\bibfnamefont {Max}\ \bibnamefont
  {Tegmark}},\ }\bibfield  {title} {\enquote {\bibinfo {title} {{Apparent wave
  function collapse caused by scattering}},}\ }\href {\doibase
  10.1007/BF00662807} {\bibfield  {journal} {\bibinfo  {journal} {Found. Phys.
  Lett.}\ }\textbf {\bibinfo {volume} {6}},\ \bibinfo {pages} {571} (\bibinfo
  {year} {1993})},\ \Eprint {http://arxiv.org/abs/gr-qc/9310032}
  {arXiv:gr-qc/9310032} \BibitemShut {NoStop}%
\bibitem [{\citenamefont {Giulini}\ \emph {et~al.}(1996)\citenamefont
  {Giulini}, \citenamefont {Kiefer}, \citenamefont {Joos}, \citenamefont
  {Kupsch}, \citenamefont {Stamatescu},\ and\ \citenamefont
  {Zeh}}]{Giulini:1996nw}%
  \BibitemOpen
  \bibfield  {author} {\bibinfo {author} {\bibfnamefont {D.}~\bibnamefont
  {Giulini}}, \bibinfo {author} {\bibfnamefont {C.}~\bibnamefont {Kiefer}},
  \bibinfo {author} {\bibfnamefont {E.}~\bibnamefont {Joos}}, \bibinfo {author}
  {\bibfnamefont {J.}~\bibnamefont {Kupsch}}, \bibinfo {author} {\bibfnamefont
  {I.~O.}\ \bibnamefont {Stamatescu}}, \ and\ \bibinfo {author} {\bibfnamefont
  {H.~D.}\ \bibnamefont {Zeh}},\ }\href@noop {} {\emph {\bibinfo {title}
  {{Decoherence and the appearance of a classical world in quantum theory}}}}\
  (\bibinfo {year} {1996})\BibitemShut {NoStop}%
\bibitem [{\citenamefont {Zurek}(2024)}]{Zurek:2024qfm}%
  \BibitemOpen
  \bibfield  {author} {\bibinfo {author} {\bibfnamefont {Kathryn~M.}\
  \bibnamefont {Zurek}},\ }\bibfield  {title} {\enquote {\bibinfo {title}
  {{Dark Matter Candidates of a Very Low Mass}},}\ }\href {\doibase
  10.1146/annurev-nucl-101918-023542} {\bibfield  {journal} {\bibinfo
  {journal} {Ann. Rev. Nucl. Part. Sci.}\ }\textbf {\bibinfo {volume} {74}},\
  \bibinfo {pages} {287--319} (\bibinfo {year} {2024})},\ \Eprint
  {http://arxiv.org/abs/2401.03025} {arXiv:2401.03025 [hep-ph]} \BibitemShut
  {NoStop}%
\bibitem [{\citenamefont {Badurina}\ \emph {et~al.}(2024)\citenamefont
  {Badurina}, \citenamefont {Murgui},\ and\ \citenamefont
  {Plestid}}]{Badurina:2024nge}%
  \BibitemOpen
  \bibfield  {author} {\bibinfo {author} {\bibfnamefont {Leonardo}\
  \bibnamefont {Badurina}}, \bibinfo {author} {\bibfnamefont {Clara}\
  \bibnamefont {Murgui}}, \ and\ \bibinfo {author} {\bibfnamefont {Ryan}\
  \bibnamefont {Plestid}},\ }\bibfield  {title} {\enquote {\bibinfo {title}
  {{Coherent collisional decoherence}},}\ }\href {\doibase
  10.1103/PhysRevA.110.033311} {\bibfield  {journal} {\bibinfo  {journal}
  {Phys. Rev. A}\ }\textbf {\bibinfo {volume} {110}},\ \bibinfo {pages}
  {033311} (\bibinfo {year} {2024})},\ \Eprint
  {http://arxiv.org/abs/2402.03421} {arXiv:2402.03421 [quant-ph]} \BibitemShut
  {NoStop}%
\bibitem [{\citenamefont {Bednyakov}\ and\ \citenamefont
  {Naumov}(2018)}]{Bednyakov:2018mjd}%
  \BibitemOpen
  \bibfield  {author} {\bibinfo {author} {\bibfnamefont {Vadim~A.}\
  \bibnamefont {Bednyakov}}\ and\ \bibinfo {author} {\bibfnamefont {Dmitry~V.}\
  \bibnamefont {Naumov}},\ }\bibfield  {title} {\enquote {\bibinfo {title}
  {Coherency and incoherency in neutrino-nucleus elastic and inelastic
  scattering},}\ }\href {\doibase 10.1103/PhysRevD.98.053004} {\bibfield
  {journal} {\bibinfo  {journal} {Phys. Rev. D}\ }\textbf {\bibinfo {volume}
  {98}},\ \bibinfo {pages} {053004} (\bibinfo {year} {2018})},\ \Eprint
  {http://arxiv.org/abs/1806.08768} {arXiv:1806.08768 [hep-ph]} \BibitemShut
  {NoStop}%
\bibitem [{\citenamefont {Gallis}\ and\ \citenamefont
  {Fleming}(1990{\natexlab{b}})}]{Gallis:1990}%
  \BibitemOpen
  \bibfield  {author} {\bibinfo {author} {\bibfnamefont {M.~R.}\ \bibnamefont
  {Gallis}}\ and\ \bibinfo {author} {\bibfnamefont {G.~N.}\ \bibnamefont
  {Fleming}},\ }\bibfield  {title} {\enquote {\bibinfo {title} {{Environmental
  and spontaneous localization}},}\ }\href {\doibase 10.1103/PhysRevA.42.38}
  {\bibfield  {journal} {\bibinfo  {journal} {Phys. Rev. A}\ }\textbf {\bibinfo
  {volume} {42}},\ \bibinfo {pages} {38--48} (\bibinfo {year}
  {1990}{\natexlab{b}})}\BibitemShut {NoStop}%
\bibitem [{\citenamefont {Hornberger}\ and\ \citenamefont
  {Sipe}(2003{\natexlab{b}})}]{Hornberger:2003}%
  \BibitemOpen
  \bibfield  {author} {\bibinfo {author} {\bibfnamefont {K.}~\bibnamefont
  {Hornberger}}\ and\ \bibinfo {author} {\bibfnamefont {J.~E.}\ \bibnamefont
  {Sipe}},\ }\bibfield  {title} {\enquote {\bibinfo {title} {{Collisional
  decoherence reexamined}},}\ }\href {\doibase 10.1103/PhysRevA.68.012105}
  {\bibfield  {journal} {\bibinfo  {journal} {Phys. Rev. A}\ }\textbf {\bibinfo
  {volume} {68}},\ \bibinfo {pages} {012105} (\bibinfo {year}
  {2003}{\natexlab{b}})},\ \Eprint {http://arxiv.org/abs/0303094}
  {arXiv:0303094 [quant-ph]} \BibitemShut {NoStop}%
\bibitem [{\citenamefont {Cronin}\ \emph
  {et~al.}(2009{\natexlab{a}})\citenamefont {Cronin}, \citenamefont
  {Schmiedmayer},\ and\ \citenamefont {Pritchard}}]{Cronin:2009zz}%
  \BibitemOpen
  \bibfield  {author} {\bibinfo {author} {\bibfnamefont {Alexander~D.}\
  \bibnamefont {Cronin}}, \bibinfo {author} {\bibfnamefont {Joerg}\
  \bibnamefont {Schmiedmayer}}, \ and\ \bibinfo {author} {\bibfnamefont
  {David~E.}\ \bibnamefont {Pritchard}},\ }\bibfield  {title} {\enquote
  {\bibinfo {title} {{Optics and interferometry with atoms and molecules}},}\
  }\href {\doibase 10.1103/RevModPhys.81.1051} {\bibfield  {journal} {\bibinfo
  {journal} {Rev. Mod. Phys.}\ }\textbf {\bibinfo {volume} {81}},\ \bibinfo
  {pages} {1051--1129} (\bibinfo {year} {2009}{\natexlab{a}})},\ \Eprint
  {http://arxiv.org/abs/0712.3703} {arXiv:0712.3703 [quant-ph]} \BibitemShut
  {NoStop}%
\bibitem [{\citenamefont {Rudolph}\ \emph {et~al.}(2020)\citenamefont
  {Rudolph}, \citenamefont {Wilkason}, \citenamefont {Nantel}, \citenamefont
  {Swan}, \citenamefont {Holland}, \citenamefont {Jiang}, \citenamefont
  {Garber}, \citenamefont {Carman},\ and\ \citenamefont
  {Hogan}}]{Rudolph:2019vcv}%
  \BibitemOpen
  \bibfield  {author} {\bibinfo {author} {\bibfnamefont {Jan}\ \bibnamefont
  {Rudolph}}, \bibinfo {author} {\bibfnamefont {Thomas}\ \bibnamefont
  {Wilkason}}, \bibinfo {author} {\bibfnamefont {Megan}\ \bibnamefont
  {Nantel}}, \bibinfo {author} {\bibfnamefont {Hunter}\ \bibnamefont {Swan}},
  \bibinfo {author} {\bibfnamefont {Connor~M.}\ \bibnamefont {Holland}},
  \bibinfo {author} {\bibfnamefont {Yijun}\ \bibnamefont {Jiang}}, \bibinfo
  {author} {\bibfnamefont {Benjamin~E.}\ \bibnamefont {Garber}}, \bibinfo
  {author} {\bibfnamefont {Samuel~P.}\ \bibnamefont {Carman}}, \ and\ \bibinfo
  {author} {\bibfnamefont {Jason~M.}\ \bibnamefont {Hogan}},\ }\bibfield
  {title} {\enquote {\bibinfo {title} {{Large Momentum Transfer Clock Atom
  Interferometry on the 689 nm Intercombination Line of Strontium}},}\ }\href
  {\doibase 10.1103/PhysRevLett.124.083604} {\bibfield  {journal} {\bibinfo
  {journal} {Phys. Rev. Lett.}\ }\textbf {\bibinfo {volume} {124}},\ \bibinfo
  {pages} {083604} (\bibinfo {year} {2020})},\ \Eprint
  {http://arxiv.org/abs/1910.05459} {arXiv:1910.05459 [physics.atom-ph]}
  \BibitemShut {NoStop}%
\bibitem [{\citenamefont {Baynham}\ \emph {et~al.}(2025)\citenamefont {Baynham}
  \emph {et~al.}}]{AION:2025igp}%
  \BibitemOpen
  \bibfield  {author} {\bibinfo {author} {\bibfnamefont {C.~F.~A.}\
  \bibnamefont {Baynham}} \emph {et~al.} (\bibinfo {collaboration} {AION}),\
  }\bibfield  {title} {\enquote {\bibinfo {title} {{A Prototype Atom
  Interferometer to Detect Dark Matter and Gravitational Waves}},}\ }\href@noop
  {} {\  (\bibinfo {year} {2025})},\ \Eprint {http://arxiv.org/abs/2504.09158}
  {arXiv:2504.09158 [hep-ex]} \BibitemShut {NoStop}%
\bibitem [{\citenamefont {Cronin}\ \emph
  {et~al.}(2009{\natexlab{b}})\citenamefont {Cronin}, \citenamefont
  {Schmiedmayer},\ and\ \citenamefont {Pritchard}}]{RevModPhys.81.1051}%
  \BibitemOpen
  \bibfield  {author} {\bibinfo {author} {\bibfnamefont {Alexander~D.}\
  \bibnamefont {Cronin}}, \bibinfo {author} {\bibfnamefont {J\"org}\
  \bibnamefont {Schmiedmayer}}, \ and\ \bibinfo {author} {\bibfnamefont
  {David~E.}\ \bibnamefont {Pritchard}},\ }\bibfield  {title} {\enquote
  {\bibinfo {title} {Optics and interferometry with atoms and molecules},}\
  }\href {\doibase 10.1103/RevModPhys.81.1051} {\bibfield  {journal} {\bibinfo
  {journal} {Rev. Mod. Phys.}\ }\textbf {\bibinfo {volume} {81}},\ \bibinfo
  {pages} {1051--1129} (\bibinfo {year} {2009}{\natexlab{b}})}\BibitemShut
  {NoStop}%
\bibitem [{\citenamefont {Rocco}\ \emph {et~al.}(2014)\citenamefont {Rocco},
  \citenamefont {Palmer}, \citenamefont {Valenzuela}, \citenamefont {Boyer},
  \citenamefont {Freise},\ and\ \citenamefont {Bongs}}]{Rocco_2014}%
  \BibitemOpen
  \bibfield  {author} {\bibinfo {author} {\bibfnamefont {E}~\bibnamefont
  {Rocco}}, \bibinfo {author} {\bibfnamefont {R~N}\ \bibnamefont {Palmer}},
  \bibinfo {author} {\bibfnamefont {T}~\bibnamefont {Valenzuela}}, \bibinfo
  {author} {\bibfnamefont {V}~\bibnamefont {Boyer}}, \bibinfo {author}
  {\bibfnamefont {A}~\bibnamefont {Freise}}, \ and\ \bibinfo {author}
  {\bibfnamefont {K}~\bibnamefont {Bongs}},\ }\bibfield  {title} {\enquote
  {\bibinfo {title} {Fluorescence detection at the atom shot noise limit for
  atom interferometry},}\ }\href {\doibase 10.1088/1367-2630/16/9/093046}
  {\bibfield  {journal} {\bibinfo  {journal} {New Journal of Physics}\ }\textbf
  {\bibinfo {volume} {16}},\ \bibinfo {pages} {093046} (\bibinfo {year}
  {2014})}\BibitemShut {NoStop}%
\bibitem [{\citenamefont {Lipkin}(2004)}]{lipkin2004physicsdebyewallerfactors}%
  \BibitemOpen
  \bibfield  {author} {\bibinfo {author} {\bibfnamefont {Harry~J.}\
  \bibnamefont {Lipkin}},\ }\href {https://arxiv.org/abs/cond-mat/0405023}
  {\enquote {\bibinfo {title} {Physics of debye-waller factors},}\ } (\bibinfo
  {year} {2004}),\ \Eprint {http://arxiv.org/abs/cond-mat/0405023}
  {arXiv:cond-mat/0405023 [cond-mat.mes-hall]} \BibitemShut {NoStop}%
\bibitem [{\citenamefont {Du}\ \emph {et~al.}(2023)\citenamefont {Du},
  \citenamefont {Murgui}, \citenamefont {Pardo}, \citenamefont {Wang},\ and\
  \citenamefont {Zurek}}]{Du:2023eae}%
  \BibitemOpen
  \bibfield  {author} {\bibinfo {author} {\bibfnamefont {Yufeng}\ \bibnamefont
  {Du}}, \bibinfo {author} {\bibfnamefont {Clara}\ \bibnamefont {Murgui}},
  \bibinfo {author} {\bibfnamefont {Kris}\ \bibnamefont {Pardo}}, \bibinfo
  {author} {\bibfnamefont {Yikun}\ \bibnamefont {Wang}}, \ and\ \bibinfo
  {author} {\bibfnamefont {Kathryn~M.}\ \bibnamefont {Zurek}},\ }\bibfield
  {title} {\enquote {\bibinfo {title} {{Contrast Loss from Astrophysical
  Backgrounds in Space-Based Matter-Wave Interferometers}},}\ }\href@noop {} {\
   (\bibinfo {year} {2023})},\ \Eprint {http://arxiv.org/abs/2308.02634}
  {arXiv:2308.02634 [quant-ph]} \BibitemShut {NoStop}%
\bibitem [{\citenamefont {Kaltenbaek}\ \emph {et~al.}(2016)\citenamefont
  {Kaltenbaek} \emph {et~al.}}]{Kaltenbaek:2015kha}%
  \BibitemOpen
  \bibfield  {author} {\bibinfo {author} {\bibfnamefont {Rainer}\ \bibnamefont
  {Kaltenbaek}} \emph {et~al.},\ }\bibfield  {title} {\enquote {\bibinfo
  {title} {{Macroscopic quantum resonators (MAQRO): 2015 Update}},}\ }\href
  {\doibase 10.1140/epjqt/s40507-016-0043-7} {\bibfield  {journal} {\bibinfo
  {journal} {EPJ Quant. Technol.}\ }\textbf {\bibinfo {volume} {3}},\ \bibinfo
  {pages} {5} (\bibinfo {year} {2016})},\ \Eprint
  {http://arxiv.org/abs/1503.02640} {arXiv:1503.02640 [quant-ph]} \BibitemShut
  {NoStop}%
\bibitem [{\citenamefont {Kaltenbaek}\ \emph {et~al.}(2023)\citenamefont
  {Kaltenbaek} \emph {et~al.}}]{Kaltenbaek:2023xtz}%
  \BibitemOpen
  \bibfield  {author} {\bibinfo {author} {\bibfnamefont {Rainer}\ \bibnamefont
  {Kaltenbaek}} \emph {et~al.},\ }\bibfield  {title} {\enquote {\bibinfo
  {title} {{Research campaign: Macroscopic quantum resonators (MAQRO)}},}\
  }\href {\doibase 10.1088/2058-9565/aca3cd} {\bibfield  {journal} {\bibinfo
  {journal} {Quantum Sci. Technol.}\ }\textbf {\bibinfo {volume} {8}},\
  \bibinfo {pages} {014006} (\bibinfo {year} {2023})}\BibitemShut {NoStop}%
\bibitem [{\citenamefont {Pino}\ \emph {et~al.}(2018)\citenamefont {Pino},
  \citenamefont {Prat-Camps}, \citenamefont {Sinha}, \citenamefont
  {Venkatesh},\ and\ \citenamefont {Romero-Isart}}]{Pino_2018}%
  \BibitemOpen
  \bibfield  {author} {\bibinfo {author} {\bibfnamefont {H}~\bibnamefont
  {Pino}}, \bibinfo {author} {\bibfnamefont {J}~\bibnamefont {Prat-Camps}},
  \bibinfo {author} {\bibfnamefont {K}~\bibnamefont {Sinha}}, \bibinfo {author}
  {\bibfnamefont {B~Prasanna}\ \bibnamefont {Venkatesh}}, \ and\ \bibinfo
  {author} {\bibfnamefont {O}~\bibnamefont {Romero-Isart}},\ }\bibfield
  {title} {\enquote {\bibinfo {title} {On-chip quantum interference of a
  superconducting microsphere},}\ }\href {\doibase 10.1088/2058-9565/aa9d15}
  {\bibfield  {journal} {\bibinfo  {journal} {Quantum Science and Technology}\
  }\textbf {\bibinfo {volume} {3}},\ \bibinfo {pages} {025001} (\bibinfo {year}
  {2018})}\BibitemShut {NoStop}%
\bibitem [{\citenamefont {Naraschewski}\ \emph {et~al.}(1996)\citenamefont
  {Naraschewski}, \citenamefont {Wallis}, \citenamefont {Schenzle},
  \citenamefont {Cirac},\ and\ \citenamefont {Zoller}}]{Naraschewski:1996PRA}%
  \BibitemOpen
  \bibfield  {author} {\bibinfo {author} {\bibfnamefont {M.}~\bibnamefont
  {Naraschewski}}, \bibinfo {author} {\bibfnamefont {H.}~\bibnamefont
  {Wallis}}, \bibinfo {author} {\bibfnamefont {A.}~\bibnamefont {Schenzle}},
  \bibinfo {author} {\bibfnamefont {J.~I.}\ \bibnamefont {Cirac}}, \ and\
  \bibinfo {author} {\bibfnamefont {P.}~\bibnamefont {Zoller}},\ }\bibfield
  {title} {\enquote {\bibinfo {title} {Interference of bose condensates},}\
  }\href {\doibase 10.1103/PhysRevA.54.2185} {\bibfield  {journal} {\bibinfo
  {journal} {Phys. Rev. A}\ }\textbf {\bibinfo {volume} {54}},\ \bibinfo
  {pages} {2185--2196} (\bibinfo {year} {1996})}\BibitemShut {NoStop}%
\bibitem [{\citenamefont {Asenbaum}\ \emph {et~al.}(2020)\citenamefont
  {Asenbaum}, \citenamefont {Overstreet}, \citenamefont {Kim}, \citenamefont
  {Curti},\ and\ \citenamefont {Kasevich}}]{Asenbaum:2020era}%
  \BibitemOpen
  \bibfield  {author} {\bibinfo {author} {\bibfnamefont {Peter}\ \bibnamefont
  {Asenbaum}}, \bibinfo {author} {\bibfnamefont {Chris}\ \bibnamefont
  {Overstreet}}, \bibinfo {author} {\bibfnamefont {Minjeong}\ \bibnamefont
  {Kim}}, \bibinfo {author} {\bibfnamefont {Joseph}\ \bibnamefont {Curti}}, \
  and\ \bibinfo {author} {\bibfnamefont {Mark~A.}\ \bibnamefont {Kasevich}},\
  }\bibfield  {title} {\enquote {\bibinfo {title} {{Atom-Interferometric Test
  of the Equivalence Principle at the $10^{-12}$ Level}},}\ }\href {\doibase
  10.1103/PhysRevLett.125.191101} {\bibfield  {journal} {\bibinfo  {journal}
  {Phys. Rev. Lett.}\ }\textbf {\bibinfo {volume} {125}},\ \bibinfo {pages}
  {191101} (\bibinfo {year} {2020})},\ \Eprint
  {http://arxiv.org/abs/2005.11624} {arXiv:2005.11624 [physics.atom-ph]}
  \BibitemShut {NoStop}%
\bibitem [{\citenamefont {Overstreet}\ \emph {et~al.}(2022)\citenamefont
  {Overstreet}, \citenamefont {Asenbaum}, \citenamefont {Curti}, \citenamefont
  {Kim},\ and\ \citenamefont {Kasevich}}]{Overstreet:2022}%
  \BibitemOpen
  \bibfield  {author} {\bibinfo {author} {\bibfnamefont {C.}~\bibnamefont
  {Overstreet}}, \bibinfo {author} {\bibfnamefont {P.}~\bibnamefont
  {Asenbaum}}, \bibinfo {author} {\bibfnamefont {J.}~\bibnamefont {Curti}},
  \bibinfo {author} {\bibfnamefont {M.}~\bibnamefont {Kim}}, \ and\ \bibinfo
  {author} {\bibfnamefont {M.~A.}\ \bibnamefont {Kasevich}},\ }\bibfield
  {title} {\enquote {\bibinfo {title} {{Observation of a gravitational
  aharonov- bohm effect}},}\ }\href {\doibase doi: 10.1126/science.abl7152}
  {\bibfield  {journal} {\bibinfo  {journal} {Science}\ }\textbf {\bibinfo
  {volume} {375}},\ \bibinfo {pages} {226–229} (\bibinfo {year}
  {2022})}\BibitemShut {NoStop}%
\bibitem [{\citenamefont {{ams OSRAM}}()}]{IRlamp}%
  \BibitemOpen
  \bibfield  {author} {\bibinfo {author} {\bibnamefont {{ams OSRAM}}},\ }\href
  {https://docs.rs-online.com/0c0a/0900766b81545a1d.pdf} {\enquote {\bibinfo
  {title} {Oslon black series (850 nm) – 90°: Sfh 4715a datasheet},}\
  }\bibinfo {note} {Accessed 4 Aug 2025}\BibitemShut {NoStop}%
\bibitem [{\citenamefont {Gregoire}\ \emph {et~al.}(2015)\citenamefont
  {Gregoire}, \citenamefont {Hromada}, \citenamefont {Holmgren}, \citenamefont
  {Trubko},\ and\ \citenamefont {Cronin}}]{Gregoire_2015}%
  \BibitemOpen
  \bibfield  {author} {\bibinfo {author} {\bibfnamefont {Maxwell~D.}\
  \bibnamefont {Gregoire}}, \bibinfo {author} {\bibfnamefont {Ivan}\
  \bibnamefont {Hromada}}, \bibinfo {author} {\bibfnamefont {William~F.}\
  \bibnamefont {Holmgren}}, \bibinfo {author} {\bibfnamefont {Raisa}\
  \bibnamefont {Trubko}}, \ and\ \bibinfo {author} {\bibfnamefont
  {Alexander~D.}\ \bibnamefont {Cronin}},\ }\bibfield  {title} {\enquote
  {\bibinfo {title} {Measurements of the ground-state polarizabilities of cs,
  rb, and k using atom interferometry},}\ }\href {\doibase
  10.1103/physreva.92.052513} {\bibfield  {journal} {\bibinfo  {journal}
  {Physical Review A}\ }\textbf {\bibinfo {volume} {92}} (\bibinfo {year}
  {2015}),\ 10.1103/physreva.92.052513}\BibitemShut {NoStop}%
\bibitem [{\citenamefont {Physics}(2025)}]{kimball_electron_gun_systems}%
  \BibitemOpen
  \bibfield  {author} {\bibinfo {author} {\bibfnamefont {Kimball}\ \bibnamefont
  {Physics}},\ }\href
  {https://www.kimballphysics.com/learning_center/electron-gun-beam-systems/}
  {\enquote {\bibinfo {title} {Electron gun (beam) systems},}\ } (\bibinfo
  {year} {2025}),\ \bibinfo {note} {accessed: 2025-06-05}\BibitemShut {NoStop}%
\bibitem [{\citenamefont {Gien}(1991)}]{PhysRevA.44.5693}%
  \BibitemOpen
  \bibfield  {author} {\bibinfo {author} {\bibfnamefont {T.~T.}\ \bibnamefont
  {Gien}},\ }\bibfield  {title} {\enquote {\bibinfo {title} {Elastic scattering
  of electrons and positrons from rubidium},}\ }\href {\doibase
  10.1103/PhysRevA.44.5693} {\bibfield  {journal} {\bibinfo  {journal} {Phys.
  Rev. A}\ }\textbf {\bibinfo {volume} {44}},\ \bibinfo {pages} {5693--5701}
  (\bibinfo {year} {1991})}\BibitemShut {NoStop}%
\bibitem [{\citenamefont {Letokhov}(2007)}]{letokhov2007laser}%
  \BibitemOpen
  \bibfield  {author} {\bibinfo {author} {\bibfnamefont {V.}~\bibnamefont
  {Letokhov}},\ }\href {https://books.google.ch/books?id=6QlREAAAQBAJ} {\emph
  {\bibinfo {title} {Laser Control of Atoms and Molecules}}},\ International
  series of monographs on physics\ (\bibinfo  {publisher} {OUP Oxford},\
  \bibinfo {year} {2007})\BibitemShut {NoStop}%
\bibitem [{\citenamefont {Cheong}\ \emph {et~al.}(2025)\citenamefont {Cheong},
  \citenamefont {Rodd},\ and\ \citenamefont {Wang}}]{Cheong:2024ose}%
  \BibitemOpen
  \bibfield  {author} {\bibinfo {author} {\bibfnamefont {Dhong~Yeon}\
  \bibnamefont {Cheong}}, \bibinfo {author} {\bibfnamefont {Nicholas~L.}\
  \bibnamefont {Rodd}}, \ and\ \bibinfo {author} {\bibfnamefont {Lian-Tao}\
  \bibnamefont {Wang}},\ }\bibfield  {title} {\enquote {\bibinfo {title}
  {{Quantum description of wave dark matter}},}\ }\href {\doibase
  10.1103/PhysRevD.111.015028} {\bibfield  {journal} {\bibinfo  {journal}
  {Phys. Rev. D}\ }\textbf {\bibinfo {volume} {111}},\ \bibinfo {pages}
  {015028} (\bibinfo {year} {2025})},\ \Eprint
  {http://arxiv.org/abs/2408.04696} {arXiv:2408.04696 [hep-ph]} \BibitemShut
  {NoStop}%
\bibitem [{\citenamefont {Knapen}\ \emph {et~al.}(2017)\citenamefont {Knapen},
  \citenamefont {Lin},\ and\ \citenamefont {Zurek}}]{Knapen:2017xzo}%
  \BibitemOpen
  \bibfield  {author} {\bibinfo {author} {\bibfnamefont {Simon}\ \bibnamefont
  {Knapen}}, \bibinfo {author} {\bibfnamefont {Tongyan}\ \bibnamefont {Lin}}, \
  and\ \bibinfo {author} {\bibfnamefont {Kathryn~M.}\ \bibnamefont {Zurek}},\
  }\bibfield  {title} {\enquote {\bibinfo {title} {{Light Dark Matter: Models
  and Constraints}},}\ }\href {\doibase 10.1103/PhysRevD.96.115021} {\bibfield
  {journal} {\bibinfo  {journal} {Phys. Rev. D}\ }\textbf {\bibinfo {volume}
  {96}},\ \bibinfo {pages} {115021} (\bibinfo {year} {2017})},\ \Eprint
  {http://arxiv.org/abs/1709.07882} {arXiv:1709.07882 [hep-ph]} \BibitemShut
  {NoStop}%
\bibitem [{\citenamefont {Baxter}\ \emph {et~al.}(2021)\citenamefont {Baxter}
  \emph {et~al.}}]{Baxter:2021pqo}%
  \BibitemOpen
  \bibfield  {author} {\bibinfo {author} {\bibfnamefont {D.}~\bibnamefont
  {Baxter}} \emph {et~al.},\ }\bibfield  {title} {\enquote {\bibinfo {title}
  {{Recommended conventions for reporting results from direct dark matter
  searches}},}\ }\href {\doibase 10.1140/epjc/s10052-021-09655-y} {\bibfield
  {journal} {\bibinfo  {journal} {Eur. Phys. J. C}\ }\textbf {\bibinfo {volume}
  {81}},\ \bibinfo {pages} {907} (\bibinfo {year} {2021})},\ \Eprint
  {http://arxiv.org/abs/2105.00599} {arXiv:2105.00599 [hep-ex]} \BibitemShut
  {NoStop}%
\bibitem [{\citenamefont {Parker}\ \emph {et~al.}(2018)\citenamefont {Parker},
  \citenamefont {Yu}, \citenamefont {Zhong}, \citenamefont {Estey},\ and\
  \citenamefont {M{\"u}ller}}]{Parker:2018vye}%
  \BibitemOpen
  \bibfield  {author} {\bibinfo {author} {\bibfnamefont {Richard~H.}\
  \bibnamefont {Parker}}, \bibinfo {author} {\bibfnamefont {Chenghui}\
  \bibnamefont {Yu}}, \bibinfo {author} {\bibfnamefont {Weicheng}\ \bibnamefont
  {Zhong}}, \bibinfo {author} {\bibfnamefont {Brian}\ \bibnamefont {Estey}}, \
  and\ \bibinfo {author} {\bibfnamefont {Holger}\ \bibnamefont {M{\"u}ller}},\
  }\bibfield  {title} {\enquote {\bibinfo {title} {{Measurement of the
  fine-structure constant as a test of the Standard Model}},}\ }\href {\doibase
  10.1126/science.aap7706} {\bibfield  {journal} {\bibinfo  {journal}
  {Science}\ }\textbf {\bibinfo {volume} {360}},\ \bibinfo {pages} {191}
  (\bibinfo {year} {2018})},\ \Eprint {http://arxiv.org/abs/1812.04130}
  {arXiv:1812.04130 [physics.atom-ph]} \BibitemShut {NoStop}%
\bibitem [{\citenamefont {El-Neaj}\ \emph {et~al.}(2020)\citenamefont {El-Neaj}
  \emph {et~al.}}]{AEDGE:2019nxb}%
  \BibitemOpen
  \bibfield  {author} {\bibinfo {author} {\bibfnamefont {Yousef~Abou}\
  \bibnamefont {El-Neaj}} \emph {et~al.} (\bibinfo {collaboration} {AEDGE}),\
  }\bibfield  {title} {\enquote {\bibinfo {title} {{AEDGE: Atomic Experiment
  for Dark Matter and Gravity Exploration in Space}},}\ }\href {\doibase
  10.1140/epjqt/s40507-020-0080-0} {\bibfield  {journal} {\bibinfo  {journal}
  {EPJ Quant. Technol.}\ }\textbf {\bibinfo {volume} {7}},\ \bibinfo {pages}
  {6} (\bibinfo {year} {2020})},\ \Eprint {http://arxiv.org/abs/1908.00802}
  {arXiv:1908.00802 [gr-qc]} \BibitemShut {NoStop}%
\bibitem [{\citenamefont {Murata}\ and\ \citenamefont
  {Tanaka}(2015)}]{Murata:2014nra}%
  \BibitemOpen
  \bibfield  {author} {\bibinfo {author} {\bibfnamefont {Jiro}\ \bibnamefont
  {Murata}}\ and\ \bibinfo {author} {\bibfnamefont {Saki}\ \bibnamefont
  {Tanaka}},\ }\bibfield  {title} {\enquote {\bibinfo {title} {{A review of
  short-range gravity experiments in the LHC era}},}\ }\href {\doibase
  10.1088/0264-9381/32/3/033001} {\bibfield  {journal} {\bibinfo  {journal}
  {Class. Quant. Grav.}\ }\textbf {\bibinfo {volume} {32}},\ \bibinfo {pages}
  {033001} (\bibinfo {year} {2015})},\ \Eprint {http://arxiv.org/abs/1408.3588}
  {arXiv:1408.3588 [hep-ex]} \BibitemShut {NoStop}%
\bibitem [{\citenamefont {Hardy}\ and\ \citenamefont
  {Lasenby}(2017)}]{Hardy:2016kme}%
  \BibitemOpen
  \bibfield  {author} {\bibinfo {author} {\bibfnamefont {Edward}\ \bibnamefont
  {Hardy}}\ and\ \bibinfo {author} {\bibfnamefont {Robert}\ \bibnamefont
  {Lasenby}},\ }\bibfield  {title} {\enquote {\bibinfo {title} {{Stellar
  cooling bounds on new light particles: plasma mixing effects}},}\ }\href
  {\doibase 10.1007/JHEP02(2017)033} {\bibfield  {journal} {\bibinfo  {journal}
  {JHEP}\ }\textbf {\bibinfo {volume} {02}},\ \bibinfo {pages} {033} (\bibinfo
  {year} {2017})},\ \Eprint {http://arxiv.org/abs/1611.05852} {arXiv:1611.05852
  [hep-ph]} \BibitemShut {NoStop}%
\bibitem [{\citenamefont {Tulin}\ and\ \citenamefont
  {Yu}(2018)}]{Tulin:2017ara}%
  \BibitemOpen
  \bibfield  {author} {\bibinfo {author} {\bibfnamefont {Sean}\ \bibnamefont
  {Tulin}}\ and\ \bibinfo {author} {\bibfnamefont {Hai-Bo}\ \bibnamefont
  {Yu}},\ }\bibfield  {title} {\enquote {\bibinfo {title} {{Dark Matter
  Self-interactions and Small Scale Structure}},}\ }\href {\doibase
  10.1016/j.physrep.2017.11.004} {\bibfield  {journal} {\bibinfo  {journal}
  {Phys. Rept.}\ }\textbf {\bibinfo {volume} {730}},\ \bibinfo {pages} {1--57}
  (\bibinfo {year} {2018})},\ \Eprint {http://arxiv.org/abs/1705.02358}
  {arXiv:1705.02358 [hep-ph]} \BibitemShut {NoStop}%
\end{thebibliography}%

\end{document}